%% file: NIT.tex
\documentclass[12pt]{article}
\usepackage{amsmath,bbm,amssymb,amsthm}
\usepackage{graphicx,bm,dsfont}
\usepackage{enumerate,mathtools}
\usepackage{comment,caption,multirow}
\usepackage{natbib}
\usepackage{subcaption,bigstrut}
\usepackage{url} 
\usepackage{booktabs,float}
\usepackage[colorlinks,citecolor=blue,urlcolor=blue,linkcolor = blue]{hyperref}
\usepackage{algorithm}

\theoremstyle{thmstyleone}%
\newtheorem{theorem}{Theorem}
\newtheorem{proposition}{Proposition}%
\theoremstyle{thmstyletwo}%
\newtheorem{example}{Example}%
\newtheorem{remark}{Remark}%
\newtheorem{corollary}{Corollary}
\newtheorem{lemma}{Lemma}
\newtheorem{definition}{Definition}

\theoremstyle{thmstylethree}%

\newcommand{\beq}{\begin{equation}}
	\newcommand{\eeq}{\end{equation}}
\newcommand{\beas}{\begin{eqnarray*}}
	\newcommand{\eeas}{\end{eqnarray*}}
\newcommand{\bea}{\begin{eqnarray}}
	\newcommand{\eea}{\end{eqnarray}}
\newcommand{\bei}{\begin{itemize}}
	\newcommand{\eei}{\end{itemize}}
\newcommand{\ben}{\begin{enumerate}}
	\newcommand{\een}{\end{enumerate}}
\newcommand{\bet}{\begin{theorem}}
	\newcommand{\eet}{\end{theorem}}
\newcommand{\bel}{\begin{lemma}}
	\newcommand{\eel}{\end{lemma}}
\newcommand{\bep}{\begin{proposition}}
	\newcommand{\eep}{\end{proposition}}
\newcommand{\bed}{\begin{definition}}
	\newcommand{\eed}{\end{definition}}
\newcommand{\bec}{\begin{corollary}}
	\newcommand{\eec}{\end{corollary}}
\newcommand{\bex}{\begin{example}}
	\newcommand{\eex}{\end{example}}
\newcommand{\E}{\mathbb{E}}
\newcommand{\R}{\mathbb{R}}
\newcommand{\cN}{\mathcal{N}}
\newcommand{\red}[1]{{\textcolor{black}{#1}}}	
\DeclarePairedDelimiter{\norm}{\lVert}{\rVert}
\input{macros.tex}

\def\sf{{\cal F}}

\newcommand{\blind}{1}

\begin{document}
	
\def\spacingset#1{\renewcommand{\baselinestretch}%
	{#1}\small\normalsize} \spacingset{1}


\if1\blind
{
	\title{\bf Empirical Bayes Estimation with Side Information: A Nonparametric
		Integrative Tweedie Approach}
	\author{Jiajun Luo$^1$, Trambak Banerjee$^2$, Gourab Mukherjee$^1$ and Wenguang Sun$^3$\hspace{.5cm}\\[1ex]
		University of Southern California$^1$, University of Kansas$^2$ and Zhejiang University$^3$\\
	}
	\date{}
	\maketitle
} \fi

\if0\blind
{
	\bigskip
	\bigskip
	\bigskip
	\begin{center}
		{\LARGE\bf Harnessing The Collective Wisdom: Integrative False Discovery Rate Control Using Binary Decision Sequences From Diverse Sources}
	\end{center}
} \fi
\vspace{-30pt}
\begin{abstract}
We investigate the problem of compound estimation of normal means while accounting for the presence of side information. Leveraging the empirical Bayes framework, we develop a nonparametric integrative Tweedie (NIT) approach that incorporates structural knowledge encoded in multivariate auxiliary data to enhance the precision of compound estimation. Our approach employs convex optimization tools to estimate the gradient of the log-density directly, enabling the incorporation of structural constraints. We conduct theoretical analyses of the asymptotic risk of NIT and establish the rate at which NIT converges to the oracle estimator. As the dimension of the auxiliary data increases, we accurately quantify the improvements in estimation risk and the associated deterioration in convergence rate. The numerical performance of NIT is illustrated through the analysis of both simulated and real data, demonstrating its superiority over existing methods. 
\end{abstract}
\noindent%
\spacingset{1.25} 
\section{Introduction} 
In data-intensive fields, such as genomics, neuroimaging, and signal processing, vast amounts of data are collected, often accompanied by various types of side information. We consider a compound estimation problem where $\bm Y=(Y_i: 1\leq i\leq n)$ is a vector of summary statistics and serves as the primary data for analysis. In addition, we collect $K$ auxiliary sequences $\bm{S}^{(k)}=(S^{(k)}_{i}: 1\leq i\leq n)$, $1\le k \le K$, alongside the primary data. Suppose the elements in $\bm Y$ follow normal distributions 
\begin{equation}\label{model1}
Y_i=\theta_i+\epsilon_i, \quad  \epsilon_i\sim N(0, \sigma^2), \quad 1\leq i\leq n,
\end{equation}
where $\theta_i=\E(Y_i)$ represents the true underlying effect size for the $i$th study unit. {Following \cite{BroGre09,Efr11,ignatiadis2019empirical} we assume that $\sigma^2$ is known or can be well estimated from the data. For instance, in practical applications we often observe replicates for some of the observations using which $\sigma^2$ can be consistently estimated. Moreover, if we are in a rapid trend changing environments where the variances are stationary then $\sigma^2$ can be estimated from past data (see, for instance, Section 2.4 of \cite{banerjee2021improved}).} Let $\bm{S}_i= (S^1_{i}, \cdots, S^K_i)^T$ denote the side information associated with unit $i$ and $\bm{S} = (\bm{S}_1, \cdots, \bm{S}_n)^T$ the auxiliary data matrix. Assume that $\bm S_i$ follow some unspecified multivariate distribution $F_S$. Our task is to estimate the high-dimensional parameter $\bm\theta=\{\theta_i: 1\leq i\leq n\}$ given both primary and auxiliary data. 

{Conventional meta-analytical methods frequently encounter two limitations. Firstly, these methods often assign equal importance to the primary and auxiliary data, relying on weighting strategies to calculate an overall effect by integrating data from multiple sources. Nevertheless, this approach may result in biased estimates of $\theta_i$ when the distributions of $\bm Y$ and $\bm S^{(k)}$ differ.} Secondly, conventional techniques, which are designed to handle a small number of parameters, can become highly inefficient for large-scale estimation problems. This inefficiency is particularly pronounced when $\bm\theta$ is in high dimensions, where valuable structural knowledge of $\bm\theta$ can be extracted from both primary and auxiliary data and exploited to construct more efficient inference procedures. 

This article presents an empirical Bayes approach to integrative compound estimation with side information.  The framework provides a flexible and powerful tool that can effectively integrate information from multiple sources. {Our method capitalizes on the structural knowledge present in auxiliary data, which can be highly informative and has the potential to greatly enhance estimation accuracy when properly assimilated into the decision-making process.} In what follows, we begin by presenting an overview of the progress made in this research direction and identify relevant issues. This will be followed by an exposition of our methodology for addressing the challenges. Finally, we discuss related works and highlight our contributions. 

\subsection{Compound decisions, structural knowledge and side information}
Consider a compound decision problem where we make simultaneous inference of $n$ parameters 
$(\theta_i: 1\leq i\leq n)$ based on summary statistics $(Y_i: 1\leq i\leq n)$ from $n$ independent experiments. Let $\bm{\delta}=(\delta_i: 1\leq i\leq n)$ be the decision rule, i.e., our estimate of $\theta_i$. Several classical ideas, such as the compound decision theory \citep{Rob51}, empirical Bayes (EB) methods \citep{Rob64}, and James-Stein shrinkage estimator \citep{Ste56}, alongside more recent multiple testing methodologies \citep{Efr01, SunCai07}, have demonstrated that structural information of the data can be leveraged to construct more efficient classification, estimation, and multiple testing procedures. For example, the subminimax rule in \cite{Rob51} has shown that the disparity in the proportions of positive and negative signals can be incorporated into inference to reduce the misclassification rate. Similarly, the adaptive $z$-value procedure in \cite{SunCai07} has demonstrated that the shape of the alternative distribution can be utilized to construct more powerful false discovery rate (FDR, \citealp{BenHoc95}) procedures.

When auxiliary data is taken into account, the inference units become heterogeneous. This heterogeneity provides new structural knowledge that can be leveraged to further improve the efficiency of existing methods. For instance, in genomics research, prior data and domain knowledge can be used to define a prioritized subset of genes. \cite{RoeWas09} proposed to up-weight the $p$-values in prioritized subsets where genes are more likely to be associated with the disease. Structured multiple testing is a crucial area which has garnered considerable attention. A partial list of references, including \cite{LeiFit18, Caietal19, LiBar19, IgnHub20, RenCan20}, demonstrates that the power of existing FDR methods can be substantially improved by utilizing auxiliary data to assign differential weights or to set varied thresholds to corresponding test statistics. Similar ideas have been adopted by some recent works on shrinkage estimation. For instance, \cite{Weietal18} and \cite{Banetal19} propose to incorporate side information into inference by first creating groups, then constructing group-wise linear shrinkage or soft-thresholding estimators.


\subsection{Nonparametric integrative Tweedie}

{Tweedie's formula is an elegant and celebrated result that has received renewed interests recently \citep{JiaZha09, BroGre09, Efr11, KoeMiz14,ignatiadis2019empirical,SahGun20,kim2022flexible,zhang2022efficient,gu2023invidious}.} Under the nonparametric empirical Bayes framework, the formula is particularly appealing for large-scale estimation problems for it is simple to implement, {removes selection bias \citep{Efr11} and enjoys frequentist optimality properties \citep{Xieetal12}}. 

The EB implementation of Tweedie's formula has been extensively studied in the literature.  \cite{Zha97} demonstrated that a truncated generalized empirical Bayes (GEB) estimator asymptotically achieves both Bayes and minimax risks. Additionally, the non-parametric maximum likelihood estimate (NPMLE, \citealp{KieWof56}) approach and the broader class of g-modeling approaches \citep{Efr16,shen2022empirical} implement Tweedie's formula by estimating the unknown prior distribution $G$ through the Kiefer-Wolfwitz estimator \citep{JiaZha09}. {The NPMLE approach enjoys desirable asymptotic optimality properties in a wide range of problems \citep{jana2023empirical,jiang2010empirical,soloff2021multivariate,polyanskiy2020self}. In contrast to the NPMLE, \cite{BroGre09} proposed the f-modeling strategy, which implements Tweedie's formula directly by estimating the marginal density of observations using Gaussian kernels.} This nonparametric EB estimator achieves asymptotic optimality in both dense and sparse regimes. Empirically, NPMLE outperforms the kernel method by \cite{BroGre09}. However, the algorithm for NPMLE by \cite{JiaZha09} cannot handle data-intensive applications due to its computational complexity. The connection between compound estimation and convex optimization was established by \cite{KoeMiz14}, which casts NPMLE as a convex program, resulting in fast and stable algorithms that outperform competing methods; see \cite{GuKoe17}, \cite{KoeGu17} and \cite{SahGun20} for recent works in this direction. {However, in the context of the g-modeling strategy, direct non-parametric assimilation of covariates has not been thoroughly investigated, and these approaches can exhibit significant computational complexity, particularly when dealing with covariates of moderate to high dimensions.}

To effectively extract and incorporate useful structural information from both primary and auxiliary data, we propose a nonparametric integrative Tweedie (NIT) approach to compound estimation of normal means. NIT utilizes the f-modeling strategy, which involves directly estimating the log-gradient of the conditional distribution of $Y$ given $\bm{S}$, {also known as the score function}, thereby eliminating the need for a deconvolution estimator for {the unknown mixing distribution}. 
We recast compound estimation via NIT as a convex program using a well-designed reproducing kernel Hilbert space (RKHS) representation of Stein's discrepancy. By searching for feasible score embeddings in the RKHS, we obtain the optimal shrinkage factor, resulting in a computationally efficient and scalable algorithm that exhibits superior empirical performance, even in high-dimensional covariate settings. The kernelized optimization framework also provides a rigorous and powerful mathematical interface for theoretical analysis. Leveraging the RKHS theory and concentration theories of V-statistics, we derive the approximate order of the kernel bandwidth, establish the asymptotic optimality of the data-driven NIT procedure, and explicitly characterize the impact of covariate dimension on the rate of convergence.

{In recent years, the theoretical foundations of score estimation approaches—particularly in the context of diffusion models used in generative AI for image generation—have garnered significant attention (see, for instance,\cite{wibisono2024optimal,zhang2024minimax,dou2024optimal} and the references therein). While these methods are primarily designed for diffusion models, their connections to our approach for estimating the score function are mainly theoretical. 
	In particular, we note that the theoretical convergence rates established in our paper (Section \ref{sec.theory}) are weaker than those achieved for score estimation in \citet{wibisono2024optimal,zhang2024minimax}. However, like these works, we also capture the exponential decay in convergence rates as the dimension $K$ of the side information vector $\bm S_i$ increases (see Theorem \ref{thm:empirical_bound}). Furthermore, both \citet{wibisono2024optimal} and \citet{zhang2024minimax} rely on ratio estimators to learn the score function and are similar in spirit to the approach pursued in \citet{BroGre09}. The numerical experiments in Section \ref{simu.sec} demonstrate superior performance of our proposed method compared to such ratio estimators across various regimes.}
%
\subsection{Our contributions}
\noindent {\bf Methodological contributions.} NIT offers several advantages over existing shrinkage estimators. Firstly, it provides a nonparametric framework for assimilating auxiliary data from multiple sources, setting it apart from existing works such as \citep{Keetal14, Cohetal13, KouYan17,IgnWag19}. Unlike these methods, NIT does not require the specification of any functional relationship, and its asymptotic optimality holds for a wider class of prior distributions. Secondly, NIT has the ability to incorporate various types of side information and effectively handle multivariate covariates. By contrast, \cite{Weietal18, Banetal19} only focus on the variance or sparsity structure, and both methods can only deal with univariate covariates by adopting a grouping approach. However, under the multivariate covariate setting, it may be infeasible to determine the optimal number of groups and to search for the ideal grouping structure. Furthermore, grouping involves discretizing a continuous variable, leading to a loss of efficiency. 
Finally, NIT is a fast, scalable, and flexible tool that can incorporate various structural constraints and produce stable estimates. 

\smallskip 
\noindent {\bf Theoretical contributions.} Firstly, we establish the convergence rates of NIT to the oracle integrative Tweedie estimator, which explicitly characterizes the improvements in estimation risk by leveraging auxiliary data. Secondly, our theoretical analysis precisely quantifies the deterioration in convergence rates as the dimension of the side information increases, providing important caveats on utilizing high-dimensional auxiliary data. 
For this theoretical analysis, we introduce new analytical tools that formalize the $L_p$ risk properties of Kernelized Stein Discrepancy (KSD) based estimators. To rigorously prove results for the $L_p$ risk, we establish a local isometry between the $L_p$ risk and the RKHS norm of the proposed estimator. Related KSD-based works \citep{Liuetal16-KSD,Chwetal16,Banetal19-DLE} assume the existence of such local isometries without providing a formal analysis. The probability tools developed here can be of independent interest for decision theorists; particularly our techniques have demonstrated their usefulness for analyzing the $L_p$ risk of recently proposed KSD methods, in the context of both heteroskedastic normal means problem \citep{Banetal21-nest} and mixed effects models \citep{Banerjee2023}. {For instance, in the context of the heteroskedastic normal means problem, \cite{Banetal21-nest} (BFJMS24) consider the following hierarchical model: \(Y_i | (\theta_i,\sigma_i^2) \stackrel{ind.}{\sim} N(\theta_i, \sigma_i^2),~\theta_i\mid\sigma_i\stackrel{ind.}{\sim}G_\mu(\cdot|\sigma_i),~\sigma_i\stackrel{i.i.d.}{\sim}H_\sigma(\cdot)\), where $G_\mu(\cdot\mid\sigma_i)$ and $H_\sigma(\cdot)$ are unspecified prior distributions, and develop novel techniques to address non-exchangeability in coordinate-wise rules arising from heterogeneous variances. We note that this setting does not fall within the scope of the homoskedastic framework considered in our paper since the models described in equations \eqref{model1} and \eqref{commonInf} assume a relationship between the location parameter \(\theta_i\) and the side information \(\bm{S}_i\), but they do not account for relationships involving variances in a heteroskedastic setup. However, the theory in our paper aligns with the theoretical derivations for Bayes-optimal rules in BFJMS 2024, and the convergence rates are related.} 

The article is organized as follows. In Section \ref{method.sec}, we discuss the empirical Bayes estimation framework, NIT estimator, and computational algorithms. Section \ref{sec.theory} delves into the theoretical properties of the NIT estimator. Sections \ref{simu.sec} and \ref{app.sec} investigate the performance of NIT using simulated and real data, respectively. Additional technical details, numerical illustrations and proofs are provided in the online Supplementary Material. {All R codes for reproducing the numerical experiments conducted in this paper are available at the following GitHub repository: \url{https://github.com/jiajunluo121/NIT}.}
\section{Methodology}\label{method.sec}
Let $\bm\delta(\bm y, \bm s)=(\delta_i: 1\leq i\leq n)$ be an estimator of $\bm \theta$, and  $\mathcal{L}^2_n (\bm{\delta}, \bm{\theta})= n^{-1}\sum_{i=1}^n ( \theta_i - \delta_i )^2$ be the corresponding loss function. We define the risk as $\R_n(\bm{\delta}, \bm{\theta}) = \E_{\bm Y,\bm S|\bm \theta} \left\{\mathcal{L}^2_n (\bm{\delta}, \bm{\theta})\right\}$ and the Bayes risk as $B_n(\bm{\delta}) = \int \R_n(\bm{\delta}, \bm{\theta})d\Pi(\bm{\theta})$, where $\Pi(\bm{\theta})$ is an unspecified prior distribution for $\bm\theta$.

We assume that the primary and auxiliary data are related through a latent vector $\bm{\xi} = (\xi_1, \cdots, \xi_n)^T$ according to the following hierarchical model:
\begin{equation}\label{commonInf}
\begin{aligned}
\theta_i &= g_{\theta} (\xi_i, \eta_{y,i}), \quad 1\leq i\leq n, \\ 
s_i^{(j)} &= g^{s}_{j}(\xi_i, \tilde \eta_{j,i}), \quad 1\le j \le K,   
\end{aligned}
\end{equation}
where $g_{\theta}$ and $g^s_j$ are unspecified functions, and $\eta_{y,i}$ and $\tilde \eta_{j,i}$ are random perturbations that are independent from $\bm\xi$. \red{
	This hierarchical model assumes that the shared information between $\theta_i$ and $\bm S_i$ is encoded by a common latent variable $\xi_i$. The relevance of the auxiliary information hinges on the noise level as well as the functional forms of $g_\theta$ and $g^s_j$. Our methodology does not require prior knowledge of $g_\theta$ and $g^s_j$, offering a versatile framework for integrating both continuous and discrete auxiliary data, and accommodates various types of side information ranging from entirely non-informative to perfectly informative. As $g_{\theta}$ and $g^s_j$ are unknown, we propose to incorporate covariate information nonparametrically. This section initially presents an oracle rule that optimally utilizes information from $\bm S$, followed by a discussion on a data-driven non-parametric rule designed to mimic the oracle rule. Subsequently, in Section~\ref{subsec:score-theory}, we delve into the frequentist risk properties of the proposed methods for a fixed sequence of $\bm \theta$.
}
{\begin{remark}
		Equations \eqref{model1} and \eqref{commonInf} can also be conceptualized as a Bayesian hierarchical model as follows:
		$$Y_i\mid(\theta_i,\bm S_i)\stackrel{ind.}{\sim} N(\theta_i,\sigma^2),~(\theta_i,\bm S_i)\mid\xi_i\stackrel{ind.}{\sim}G_\theta(\cdot\mid\xi_i)G_{\bm s}(\cdot\mid\xi_i),~\xi_i\stackrel{i.i.d.}{\sim}G_{\xi}(\cdot),
		$$
		where $G_\theta,G_{\bm s}$ and $G_\xi$ are unknown distributions. In particular, the above representation includes the hierarchical model of \cite{ignatiadis2019empirical} (see Equation 1) as a special case where, marginalizing out $\xi_i$, the conditional distribution of $\theta_i$ given $\bm S_i$ is assumed to be Gaussian.
\end{remark}}
\subsection{Oracle integrative Tweedie estimator}\label{sec.oracle}
\red{We consider an oracle with access to $f(y|\bm s)$. The oracle rule that minimizes the Bayes risk among all decision rules with the full data set, comprising both primary and auxiliary data, is referred to as the \textit{integrative Tweedie rule}. In Section~\ref{sec.theory}, we will quantify its improved performance over the Bayes rule relying solely on the primary data. The integrative Tweedie rule is presented in following proposition. It can be derived from Tweedie's formula in \citep{Efr11} and is provided in the supplement for completeness.}
\begin{proposition}[Integrative Tweedie]\label{generalizedTF}
	Consider the hierarchical model (\ref{model1}) and (\ref{commonInf}). Let $f(y|\bm{s})$ be the conditional density of $Y$ given $\bm S$ {and denote $\nabla_y\log f(y|\bm s)=\frac{\partial}{\partial y}\log f(y|\bm s)$}. The optimal estimator that minimizes the Bayes risk is $\bm{\delta}^{\pi}(\bm{y}, \bm{s}) = \left\{\delta^{\pi}(y_i, \bm s_i): 1\leq i\leq n\right\}$, where
	\begin{equation}\label{oracle}
	\delta^{\pi} (y, \bm s)= y + \sigma^2\nabla_{y} \log f(y| \bm{s}).
	\end{equation}
\end{proposition}
The integrative Tweedie rule \eqref{oracle} provides a versatile framework for integrating primary and auxiliary data. Existing literature on shrinkage estimation with side information typically requires a pre-specified form of the conditional mean function $m(\bm S_i)=E(Y_i|\bm S_i)$ \citep{Keetal14, Cohetal13, KouYan17}. In contrast, integrative Tweedie incorporates side information through a much wider class of functions $f(y|\bm{s})$ \red{(where $f$ conditioned on $\bm{s}$ is a mixture of Gaussian location densities)}, eliminating the need to pre-specify a fixed relationship between $Y$ and $\bm S$. In Section S1 of the Supplementary Material, we present two toy examples that respectively demonstrate: (a) the reduction of integrative Tweedie to an intuitive data averaging strategy when the distributions of primary and auxiliary variables match perfectly, and (b) the effectiveness of integrative Tweedie in reducing the risk (relative to ignoring $\bm S$)  even when the two distributions differ.
\subsection{Nonparametric estimation via convex programming}\label{dataDriven}
This section proposes a data-driven approach to emulate the oracle rule. Let $\mathbf {X} = (\bm x_1, \cdots, \bm x_n)^T$ denote the set of all data, where the primary sequence is denoted by $x_{1i}=y_i$ and the $(k-1)$th auxiliary sequence is denoted by $x_{ki}=s^{(k-1)}_{i}$ for $k=2,\ldots,K+1$ and $i=1,\ldots,n$. Our objective is to estimate the shrinkage factor, which is given by 
\begin{equation*}
\begin{split}
\bm{h}_f(\bm X)&=
\bigg\{\nabla_{u_1} \log f(u_1| u_2,\ldots,u_{K+1})\big\vert_{\bm{u}=\bm{x}_i}: 1\leq i\leq n\bigg\}\\
&=\{\nabla_{y_i} \log f(y_i| \bm{s}_i): 1\leq i\leq n\}.
\end{split}
\end{equation*}
We present a convex program designed to estimate $\bm{h}_f(\bm X)$. This program is motivated by the kernelized Stein's discrepancy (KSD) that we formally define in Section \ref{sec.theory}. The KSD measures the distance between a given $\bm h$ and the true score $\bm h_f$. It is always non-negative, and is equal to 0 if and only if $\bm h=\bm h_f$. Let $K_{\lambda}(\bm{x}, \bm{x}')$ be a kernel function that is integrally strictly positive definite, where $\lambda$ is a tuning parameter. A detailed discussion on the construction of kernel $K(\cdot, \cdot)$ and choice of $\lambda$ is provided in Section \ref{sec:bandwidth}. Consider the following two $n \times n$ matrices: 
$$
(\bm{K}_{\lambda})_{ij}= n^{-2}\,K_{\lambda}(\bm{x}_i, \bm{x}_j),\; (\nabla \bm{ K}_{\lambda} )_{ij}= n^{-2} \,\nabla_{x_{1j}}K_{\lambda}(\bm{x}_i, \bm{x}_j)~.
$$
Given a fixed $\lambda$, we define $\hat{\bm{h}}_{\lambda,n}$ as the solution to the following quadratic program:
\begin{equation}\label{optiprob}
\hat{\bm{h}}_{\lambda,n}=\argmin_{\bm{h} \in \bm{V}_n} \quad \bm{h}^T \bm{K}_{\lambda} \bm{h} + 2 \bm{h}^T\nabla \bm{K}_{\lambda} \bm{1}, 
\end{equation}
where $\bm{V}_n$ is a convex subset of $\mathbb{R}^n$.  Convex constraints, such as linearity and monotonicity, can be imposed through $\bm{V}_n$. Such constraints play an essential role in enhancing the stability and efficiency of compound estimation procedures \citep{KoeMiz14}; we provide a detailed discussion on these constraints in Section \ref{sec:bandwidth}. 
In Section \ref{sec.theory}, we provide theory to demonstrate that solving the convex program \eqref{optiprob} is equivalent to finding a shrinkage estimator, aided by side information, that minimizes the estimation risk.

Now combining \eqref{optiprob} and Proposition \ref{generalizedTF}, we propose the following class of nonparametric integrative Tweedie (NIT) estimators
\begin{align}\label{eq:class}
\left\{{\bm{\delta}}^{\text{NIT}}_\lambda: \lambda \in (0,\infty)\right\}, \quad \text{ where }{\bm{\delta}}^{\text{NIT}}_\lambda  = \bm{y}+{\sigma^2} \, \hat{\bm{h}}_{\lambda,n}.
\end{align}
In Section \ref{sec.theory}, we show that as $n \to \infty$ there exist choices of $\lambda$ such that the resulting estimator \eqref{eq:class} is asymptotically optimal. 

The NIT estimator \eqref{eq:class} marks a clear departure from existing NPMLE methods \citep{JiaZha09, KoeMiz14, GuKoe17}, which cannot be easily extended to handle multivariate auxiliary data. Furthermore, NIT has several additional advantages over existing empirical Bayes methods in both theory and computation. Firstly, in comparison with the NPMLE method \citep{JiaZha09}, the convex program (\ref{optiprob}) is computationally efficient and easily scalable. Secondly, \cite{BroGre09} proposed estimating the score function using the ratio $\hat f^{(1)}/\hat f$, where $\hat f$ is a kernel density estimate and $\hat f^{(1)}$ is its derivative. However, the ratio estimate can be highly unstable. In contrast, our direct optimization approach produces more stable and accurate estimates. Finally, our convex program can be fine-tuned by selecting an appropriate $\lambda$, resulting in improved numerical performance and facilitating a disciplined theoretical analysis. The criterion in \eqref{optiprob} can be rigorously analyzed to establish new rates of convergence (Sec. \ref{sec.connection}) that are previously unknown in the literature.   
\subsection{Computational details}\label{sec:bandwidth}
This section presents several computational details: (a) a discussion on how to impose convex constraints; (b) a description of how to construct kernel functions capable of handling multivariate and potentially correlated covariates; and (c) a strategy on how to select the bandwidth $\lambda$.

\noindent\textbf{\emph{1. Structural constraints.}} \red{In the convex program, we impose the constraint $\bm{1}^T \bm{h} = 0$. This constraint ensures the ``unbiasedness’’ of the estimator, as the expectation of the gradient of the log marginal density is theoretically zero. While other convex constraints can be readily integrated into the optimization, they may not be entirely appropriate. For instance, a monotonicity constraint, initially introduced in \cite{KoeMiz14}, has been shown to be highly effective in improving estimation accuracy in sequence models without auxiliary variables. To facilitate ease of presentation, assume $y_{1} \le y_{2} \le \cdots \le y_{n}$. The monotonicity constraint, expressed as $\sigma^2 h_{i-1} - \sigma^2 h_{i} \le y_{i} - y_{i-1}$ for all $i$, can be formulated as $M \bm{h} \preceq \bm{a}$ and incorporated into $\bm{V}_n$ in \eqref{optiprob} by selecting $M$ as the upper triangular matrix $M_{ij}=\sigma^2 (I\{i=j\} -I\{i=j-1\})$ and setting $\bm{a}^T=(y_{2} - y_{1},  y_{3} - y_{2},  \cdots, y_n - y_{n-1})$. However, the monotonicity constraint though satisfied by the Tweedie estimator for primary data is not satisfied by the integrative Tweedie estimator on the full data set as $y_1 \le y_2$ does not imply $\mathbb{E}(\theta_1|Y_1,\bm{S}_1)\le \mathbb{E} \theta_2|y_2,\bm{S}_2$. While our paper refrains from imposing such constraints, they can be easily integrated into the methodology if there is a preference for restricting the optimization space in \eqref{optiprob}. These additional constraints, though non-trivial, can significantly decrease the variance of the resulting estimator. Further discussion on this topic will be provided in Section 3.4.}

\noindent\textbf{\emph{ 2. Kernel functions.}} Constructing an appropriate kernel function is crucial when dealing with the complications that arise in the multivariate setting, where the auxiliary sequences may be correlated and have varying measurement units. We propose to use the Mahalanobis distance $\norm{\bm{x}- \bm{x}'}_{\Sigma_{\bm{x}}} = \sqrt{(\bm{x}- \bm{x}')^T \Sigma_{\bm{x}} ^{-1}(\bm{x}- \bm{x}')}$ in the kernel function, where $\Sigma_{\bm{x}}$ denotes the sample covariance matrix. Specifically, we employ the RBF kernel $K_{\lambda}(\bm{x}, \bm{x}') =  \exp \{-0.5\lambda^2 \norm{\bm{x} - \bm{x}'}^2_{\Sigma_{\bm{x}}} \}$, where $\lambda$ is the bandwidth parameter that can be selected through cross-validation. Mahalanobis distance is superior to the Euclidean distance as it is unit less, scale-invariant and accounts for the correlations in the data. When the auxiliary data contains both continuous and categorical variables, we propose to use the generalized Mahalanobis distance \citep{Kru87}. While we illustrate the methodology for mixed types of variables in the numerical studies, we only pursue the theory in the case where both $\bm Y$ and $\bm S$ are continuous.

\noindent\textbf{\emph{3. Modified cross-validation (MCV) for selecting $\lambda$.}}  Following \cite{Broetal13}, we propose to determine $\lambda$ via modified cross-validation (MCV). Specifically, we introduce $\eta_i \sim \cN(0,\sigma^2)$, $1\le i \le  n$, as \iid noise variables that are independent of $\bm{y}$ and $\bm{s}_1, \cdots, \bm{s}_K$. Define an uncorrelated pair $U_i = y_i + \alpha \eta_i$ and $V_i = y_i - \eta_i/\alpha$, which satisfies $\mbox{Cov}(U_i,V_i)=0$. The MCV strategy employs $\bm{U}=\{U_1, \cdots, U_n\}$ to construct estimators $\delta^{\sf IT}_\lambda(\bm{U}, \bm{S})$, while validating using $\bm{V}=(V_1, \cdots, V_n)$. Consider the following validation loss 
\begin{equation}\label{valid-loss}
\hat{L}_n(\lambda, \alpha) = \frac1n \sum_{i=1}^n \big\{\hat \delta^{\sf IT}_{\lambda,i}(\bm{U}, \bm{S})- V_i\big\}^2 - \sigma^2(1+1/ \alpha^2) ~.
\end{equation}
For small $\alpha$, $\lambda$ is determined as the value that minimizes $\hat{L}_n(\lambda, \alpha)$, i.e., 
$\hat \lambda = \argmin_{\lambda \in \Lambda}\hat{L}_n(\lambda, \alpha)
$ 
for any $\Lambda \subset \mathbb{R}^+$. The proposed NIT estimator is given by 
$
\{y_i + \sigma_y^2 \hat{\bm{h}}_{\hat \lambda,n}(i): 1\leq i\leq n\}.
$
Proposition \ref{prop.bandwidthconsist} in Section \ref{sec.ddbandwidth} establishes that the validation loss converges to the true loss, justifying the MCV algorithm.

\section{Theory}\label{sec.theory}

This section presents a large-sample theory for the data-driven NIT estimator in Equation \eqref{eq:class} {under the setting when $\sigma^2$ is known}. Our theoretical analysis focuses on the continuous case, where $\bm X_i=(Y_i, S_{i1}, \cdots, S_{ik})^T$, $i=1, \cdots, n$, are assumed to be i.i.d. samples from a continuous multivariate density $f$: $\mathbb{R}^{K+1} \to \mathbb{R}^+$. The focus on the continuous case enables an illuminating analysis of the convergence rates (Sections \ref{sec.cr.subgaussian} and \ref{sec.r_n}), providing valuable insights into the role that side information plays in compound estimation.

Let $\mathbf X=(\bm X_1, \cdots, \bm X_n)^T$ denote the $n \times (K+1)$ matrix of observations, with $\bm X^{(k)}$ being the $k$th column of $\mathbf X$. 
We define $f$ as a density function on $\mathbb{R}^{K+1}$, and $\mathfrak{h}_f(\bm{x})$ as the first conditional score (FCS) $\nabla_1 \log f(x_1|\bm{x}_{-1})$, where $\bm x_{-1}=\{x_{j}: 2\leq j\leq K+1\}$. By definition, the FCS is equivalent to the log-gradient of the joint density for the first coordinate, i.e., $\mathfrak{h}_f(\bm{x})=\nabla_1 \log f(\bm{x})$. Recall our primary objective is to estimate the parameter vector $\bm{\theta}=\E\left\{\bm X^{(1)}\right\}$. As we have demonstrated in Proposition \ref{generalizedTF}, the FCS $\mathfrak{h}_f(\bm{x})$ plays a crucial role in constructing the oracle estimator; hence accurately estimating the FCS enables us to obtain a precise estimate of $\bm{\theta}$. 

Section \ref{sec.connection} provides a detailed explanation of the relationship between compound estimation and kernelized Stein's discrepancy, and demonstrates how the FCS $\mathfrak{h}_f(\bm{x})$ can be accurately approximated by the solution to \eqref{optiprob}. The properties of the scores and convergence rates are established in Sections \ref{subsec:score-theory} to \ref{sec.r_n}, with additional results presented in Sections \ref{sec.cr.fattail} and \ref{sec.ddbandwidth}.

\subsection{Stein's discrepancy, shrinkage estimation and convex optimization} \label{sec.connection}

We begin by introducing a conditional version of the kernelized Stein's discrepancy measure, which is widely used to quantify the discrepancy between probability distributions. Specifically, we consider the Gaussian kernel function $K_{\lambda}: \mathbb{R}^{K+1}\times \mathbb{R}^{K+1} \to \mathbb{R}$ with bandwidth $\lambda$. Let $P$ and $Q$ be two distributions on $\mathbb{R}^{K+1}$, with densities denoted by $p$ and $q$, and associated first conditional score functions denoted by ${\mathfrak{h}}_p$ and ${\mathfrak{h}}_{q}$, respectively. We define the conditional kernelized Stein's discrepancy (KSD; \cite{Liuetal16-KSD, Chwetal16, Banetal19-DLE}) between $P$ and $Q$ as the kernel-weighted distance between ${\mathfrak{h}}_p$ and ${\mathfrak{h}}_{q}$, given by: 
\begin{eqnarray*}
	D_{\lambda}(P, Q) = \mathbb{E}_{(\bm{u}, \bm{v}) \stackrel{\textsf{ i.i.d. }}{\sim} P} \left\{ \mathcal{K}_{\lambda}(\bm{u}, \bm{v}) \times ({\mathfrak{h}}_p(\bm{u}) - {\mathfrak{h}}_q(\bm{u}))\times ( {\mathfrak{h}}_p(\bm{v}) - {\mathfrak{h}}_{q}(\bm{v}))\right\}.
\end{eqnarray*}
The versatility and effectiveness of KSD make it a valuable tool for many statistical problems. For example, the minimization of KSD-based criteria has been a popular technique to solve a range of  statistical problems, including new goodness-of-fit tests \citep{Liuetal16-KSD, Chwetal16,Yanetal18}, Bayesian inference \citep{LiuWan16,Oatetal17}, and simultaneous estimation \citep{Banetal19-DLE}. 

KSD is closely related to Maximum Mean Discrepancy (MMD, \cite{Greetal12}); both are measures of discrepancy between probability distributions. While KSD involves the use of kernel functions to construct an unbiased estimate of the Stein operator, MMD uses kernel functions to map the distributions into a reproducing kernel Hilbert space where the distance between their mean embeddings can be computed. Compared to MMD, KSD is particularly well-suited for empirical Bayes estimation because it can be directly constructed based on the score functions, which, as shown in Proposition \ref{generalizedTF}, yield optimal shrinkage factors. Moreover, it can be demonstrated (See Sec. 3 of \citet{jitkrittum2020testing}) that the conditional KSD defined above satisfies the following properties:
$$ 
D_{\lambda}(P, Q)\geq 0 \text{ and } D_{\lambda}(P, Q)=0 \text{ if and only if } \int \big|P(u_1|\bm{u}_{-1}) - Q(u_1|\bm{u}_{-1})\big| \, p(\bm{u}) \, d\bm{u}=0.
$$
These properties make KSD an attractive choice for testing and comparing distributions, as well as for other applications in which measuring the discrepancy between probability distributions is crucial.

We leverage the property that a value of 0 for the conditional KSD indicates the equality of the conditional distributions $P_1(\bm{u})=P(u_1|\bm{u}_{-1})$ and $Q_1(\bm{u})=Q(u_1|\bm{u}_{-1})$. However, the direct evaluation of $D_{\lambda}(P,Q)$ is challenging. To overcome this difficulty, we introduce an alternative representation of the KSD, initially introduced in \cite{Liuetal16-KSD} and \cite{Chwetal16}, that can be easily evaluated empirically. Specifically, we define a quadratic functional $\kappa_{\lambda}[{\mathfrak{h}}]$ over $\mathbb{R}^{K+1}\times \mathbb{R}^{K+1}$ for any functional ${\mathfrak{h}}:\mathbb{R}^{K+1}\to  \mathbb{R}$, given by:
\begin{equation}
\label{eq:kappa}
\kappa_{\lambda}[{\mathfrak{h}}](\bm{u}, \bm{v})= {K}_{\lambda}(\bm{u}, \bm{v}){\mathfrak{h}} (\bm{u}){\mathfrak{h}}(\bm{v}) + \nabla_{\bm{v}}{K}_{\lambda}(\bm{u}, \bm{v}) \, {\mathfrak{h}} (\bm{u})
+   \nabla_{\bm{u}}{K}_{\lambda}(\bm{u}, \bm{v})\, {\mathfrak{h}} (\bm{v})+ \nabla_{\bm{u}, \bm{v}}{K}_{\lambda}(\bm{u}, \bm{v})~.
\end{equation}
Then the KSD, which solely involves ${\mathfrak{h}}_q$, can be equivalently represented by
\beq\label{KSD-2}
D_{\lambda}(P,Q) =\mathbb{E}_{(\bm{u}, \bm{v}) \stackrel{\textsf{ i.i.d. }}{\sim} P} \{\kappa_{\lambda}[{\mathfrak{h}}_q](\bm{u},\bm{v})\}.
\eeq
We now turn to the compound estimation problem and discuss its connection to the KSD. Proposition \ref{generalizedTF} demonstrates that, when $f$ is known, the optimal estimator is constructed using $\bm{h}_f(\bm{X})=\left\{\mathfrak{h}_f(\bm{x}_1),\ldots,\mathfrak{h}_f(\bm{x}_n)\right\}^T$, the conditional score function evaluated at the $n$ observed data points. Define
\begin{align}\label{eq:S_n}
\widehat{\mathcal{S}}_{\lambda,n}(\bm h) =\bm{h}^T \bm{K}_{\lambda} \bm{h} + 2\bm{h}^T\nabla \bm{K}_{\lambda} \bm{1}+\bm{1}^T \nabla^2\bm{K}_{\lambda} \bm{1},
\end{align}
where $(\nabla^2\bm{K}_{\lambda})_{ij}=n^{-2} {\nabla_{x_{1j}}\nabla_{x_{1i}}}{K}_{\lambda}(\bm{x}_i, \bm{x}_j)$, $\bm{K}_{\lambda}$ is the $n \times n$ Gaussian kernel matrix with bandwidth $\lambda$, and $\nabla \bm{K}_{\lambda}$ is defined in Section \ref{dataDriven}. 
As the extra term $\bm{1}^T \nabla^2\bm{K}_{\lambda} \bm{1}$ is independent of $\bm h$, it is easy to see that the convex program \eqref{optiprob} is equivalent to minimizing $\widehat{\mathcal{S}}_{\lambda,n}(\bm h)$ over $\bm h$. 
When $\bm h$ is set to $\bm h_q=\left\{\mathfrak{h}_q(\bm x_1),\ldots,\mathfrak{h}_q(\bm x_n)\right\}$, \eqref{eq:S_n} becomes the empirical version of the KSD defined in \eqref{KSD-2}:
$$\widehat{\mathcal{S}}_{\lambda,n}(\bm{h}_q) = D_{\lambda}(\hat F_n,Q) = \mathbb{E}_{(\bm{u}, \bm{v}) \stackrel{\textsf{ i.i.d. }}{\sim} \hat F_n} \{\kappa_{\lambda}[{\mathfrak{h}}_q](\bm{u},\bm{v})\}  =
\frac{1}{n^2} \sum_{i,j=1}^n \kappa_{\lambda}[\mathfrak{h}_q](\bm{x}_i, \bm{x}_j)~,$$
where, \red{$\hat F_n=n^{-1}\sum_{i=1}^n \delta_{\bm{x}_i}$} is the empirical distribution function.

We will now present a heuristic explanation of the optimization criterion \eqref{optiprob}. As $n \to \infty$, $\hat F_n  \stackrel{P}{\to} F$ and it follows that $\widehat{\mathcal{S}}_{\lambda,n}({\bm{h}}_q) \stackrel{P}{\to} \mathcal{S}_\lambda({\frak{h}}_q):=D_\lambda(F,Q)$. While stronger versions of these results, such as uniform convergence, are available, we will not delve into those intricate details in this heuristic explanation of the proposed method's working principle. 
Moreover, $D_\lambda(F, Q)=0$ iff the first conditional distributions given the rest are equal, i.e., $ F(u_1|\bm{u}_{-1}) = Q(u_1|\bm{u}_{-1})$. Thus, if we could have minimized $S_{\lambda}(\mathfrak{h}_q)$ over the class $\mathcal{H}=\{\mathfrak{h}_q=\nabla_1 \log q(\bm x): q \text{ is any density on } \mathbb{R}^{K+1} \}$\footnote{For implementational ease, we relax the optimization space from the set of all conditional score functions $\mathcal{H}$  to the set all of all real functionals on $\mathbb{R}^{K+1}$. Due to the presence of structural constraints discussed in Section \ref{sec.oracle}, this relaxation has little impact on the numerical performance of the NIT estimator. Simulations in Section~\ref{simu.sec} show that the solutions to \eqref{optiprob} produce efficient estimates.}, then the minimum would be achieved at the true FCS ${\mathfrak{h}}_f$  and the minimum value would be $0$. However, $\mathcal{S}_{\lambda}({\mathfrak{h}}_q)$ involves the unknown true distribution $F$, making direct minimization impractical. Alternatively, we minimize the corresponding sample based criterion $\widehat{\mathcal{S}}_{\lambda,n}(\bm{h}_q)$ in \eqref{eq:S_n} (or equivalently, \eqref{optiprob}). In large-sample situations, we expect the sampling fluctuations to be small; hence, minimizing $\widehat{\mathcal{S}}_{\lambda,n}$ will lead to score function estimates very close to the true score functions.

\subsection{Score estimation under the $L_p$ loss} \label{subsec:score-theory}
The next three subsections formulate a rigorous theoretical framework, in the context of compound estimation, to derive the convergence rates of the proposed estimator. 

The criterion \eqref{optiprob} involves minimizing the V-statistic $\widehat{\mathcal{S}}_{\lambda,n}(\bm{h})$.  Using standard asymptotics results for V-statistics \citep{Ser09}, it follows that for any density $q$, $\widehat{\mathcal{S}}_{\lambda,n}(\bm{h}_q)\to \mathcal{S}_\lambda({\mathfrak{h}}_q)$ {in probability} as $n \to \infty$. Also, it follows from, for example, \cite{Liuetal16-KSD}, that $\hat{\bm{h}}_{\lambda,n}$, the solution to \eqref{optiprob}, satisfies: 
\begin{align}\label{eq:k-norm}
n^{-2}\sum_{i,j} K_{\lambda}(\bm{x}_i,\bm{x}_j)\left\{\hat{\bm{h}}_{\lambda,n}(i) - {\mathfrak{h}}_{f} (\bm{x}_i)\right\}\, \left\{\hat{\bm{h}}_{\lambda,n}(j) - {\mathfrak{h}}_{f} (\bm{x}_j)\right\}=O_{P}(n^{-1})
\end{align}
as $n \to \infty~$, where $\hat{\bm{h}}_{\lambda,n}(i)=\hat{\bm{h}}_{\lambda,n}(\bm{x}_i)$ for $i=1,\ldots,n$.  

While \eqref{eq:k-norm} shows that in the RKHS norm the estimates are asymptotically close to the true score functions, for most practical purposes we need to establish the convergence under the $\ell_p$ norm. For $p >0$, define $\ell_p(\hat{\bm{h}}_{\lambda,n},\mathfrak{h}_{f})=n^{-1}\sum_{i=1}^n |\hat{\bm{h}}_{\lambda,n}(\bm{x_i}) - {\mathfrak{h}}_{f} (\bm{x}_i)|^p$. The case of $p=2$ corresponds to Fisher's divergence. Denote the RKHS norm on the left side of \eqref{eq:k-norm} by $d_\lambda(\hat{\bm{h}}_{\lambda,n},\mathfrak{h}_{f})$. The essential difficulty in the analysis is that the isometry between the RKHS metric and $\ell_p$ metric may not exist. Concretely, for any $\lambda>0$, we can show that $d_\lambda\leq C_1\,\ell_2$, where $C_1$ is a constant. However,  the inequality in the other direction does not always hold. We aim to show that $\ell_2 \leq   C_2\, d_\lambda$ for some constant $C_2$; this would produce the desired bound on the $L_p$ risk. 
Next we provide an overview of the main ideas and key  contributions of the theoretical analyses in later subsections.  
\begin{figure}[!h]
	\centering
	\includegraphics[width=0.8\textwidth]{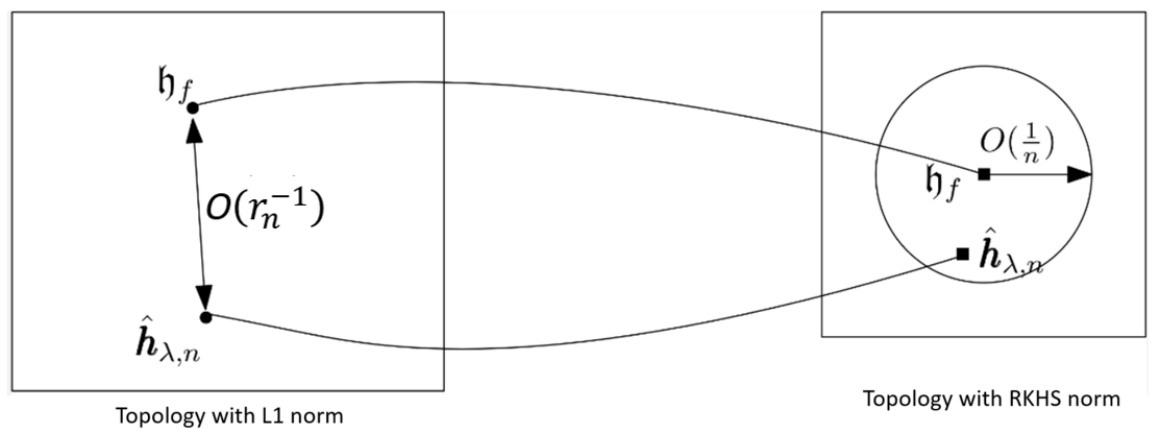}
	\caption{Schematic illustrating the relation between the RKHS and $\ell_1$ risks of $\hat{\bm{h}}_{\lambda,n}$. The (approximate) isometry can be still established. However, the error rate is increased from $n^{-1}$ to $r_n^{-1}$ due to inversion. }
\end{figure}	

In Sections \ref{sec.cr.subgaussian} and \ref{sec.cr.fattail}, we show that as $n \to \infty$ then  $\ell_2(\hat{\bm{h}}_{\lambda,n},\mathfrak{h}_{f}) \leq c_{\lambda,n}\, d_\lambda(\hat{\bm{h}}_{\lambda,n},\mathfrak{h}_{f}) \{1+o_P(1)\}$, for some $c_{\lambda,n}$ that depends on $\lambda$ and $n$ only. This result, coupled with the convergence in the RKHS norm \eqref{eq:k-norm}, produces a tolerance bound on the $L_p$ risk of  $\hat{\bm{h}}_{\lambda,n}$, which is subsequently minimized over the choice of $\lambda$ (Theorem \ref{thm:empirical_bound} in Section \ref{sec.cr.subgaussian}). However, as a result of inverting, the $n^{-1}$ error rate in \eqref{eq:k-norm} is increased to {$r_n^{-1}\approx n^{-1/(K+2)}$} for the $\ell_p$ risk (in Theorem \ref{thm:empirical_bound}, we let $p=1$). Figure \ref{fig:benefit_caveats} provides a schematic description of the phenomenon; further explanations regarding this error rate are provided after Theorem \ref{thm:empirical_bound}.    

We point out that existing KSD minimization approaches, including the proposed NIT procedure, involve first mapping the observed data into RKHS and subsequently estimating unknown quantities under the RKHS norm. A tacit assumption for developing theoretical guarantees on the $l_p$ risk is that the lower RKHS loss would also translate to lower $\ell_p$ loss; see, for example, Assumption 3 of \citet{Banetal19-DLE} and Section 5.1 of \citet{Liuetal16-KSD}. 
Heuristically, if $K_{\lambda}(\bm{u},\bm{v})=c_{\lambda} I\{\bm{u}= \bm{v}\}$, with $c_{\lambda} \to \infty$ as $n \to \infty$, then \eqref{eq:k-norm} would imply  $\ell_2(\hat{\bm{h}}_{\lambda,n},\mathfrak{h}_{f}) \to 0$. By rigorously characterizing the asymptotic quasi-geodesic between the two topologies, it can be shown that there exists such choices of $\lambda$. We provide a complete analysis of the phenomenon that our score function estimates in the RKHS transformed space has controlled $\ell_p$ risk for the compound estimation problem. This analysis, which is new in the literature, also yields the rates of convergence for the $\ell_p$ error of the proposed NIT estimator in the presence of covariates. 
\subsection{Convergence rates for sub-exponential densities}\label{sec.cr.subgaussian}
To facilitate a simpler proof, in this section we assume that the true $(K+1)$-dimensional joint density $f$ as well as its score function $\mathfrak{h}_f$ are Lipschitz continuous. We first provide  results for sub-exponential densities, which encompass the popular cases with Gaussian and exponential priors; the convergence rates for heavier-tail priors are discussed in Section \ref{sec.cr.fattail}. 

\smallskip
\noindent \textbf{Assumption 1.} The $(K+1)$ dimensional joint density $f$ is sub-exponential. 

\smallskip
\noindent Our main result is concerned with the $\ell_1$ risk of the solution from \eqref{optiprob}. The following theorem shows that the mean absolute deviation of the solution from the true score function is asymptotically negligible as $n \to \infty$. In the theorem we adopt the notation $a_n \asymp b_n$ for two sequences $a_n$ and $b_n$, which means that $c_1a_n\leq b_n \leq c_2 b_n$ for all large $n$ and some constants $c_2\geq c_1>0$.
\begin{theorem}\label{thm:empirical_bound}
	Under Assumption 1, as $n \to \infty$ with $\lambda \asymp n^{-1/(K+2)}$,
	\begin{align}\label{eq:rate-1}
	r_n \cdot \bigg(\frac{1}{n}\sum_{i=1}^n \vert \hat{\bm{h}}_{\lambda,n}(i) - \mathfrak{h}_{f}(\bm{x}_i) \vert \bigg)  \to 0 ~ \text{ in } L_1,
	\end{align}
	where, $r_n=n^{1/(K+2)}(\log (n))^{-(2K+5)}$. 
\end{theorem}
\begin{remark}
	It follows immediately from Theorem \ref{thm:empirical_bound} that the deviations between our proposed estimate and the oracle estimator in (\ref{oracle}) obeys:
	\begin{align}\label{eq:rate-2}
	r_n\left(\frac{1}{n}\sum_{i=1}^n \left|\hat{\bm{\delta}}_{\lambda}^{\sf IT}(i)-{\delta}^{\pi}_i(y_i|\bm{s}_i)\right| \right)  \to 0 \text{ in } L_1 \text{ as } n \to \infty.
	\end{align}
	\red{Under the classical setting with no auxiliary data $(K=0)$, we achieve the traditional $\sqrt{n}-$rate as established in \citet{JiaZha09}. However, the convergence rate $r_n\sim n^{1/(K+2)}$ (barring poly-log terms)  decreases polynomially in $n$ as $K$ increases.
		Noting that, for a $(K+1)$ dimensional Gaussian density, the rate of convergence of the mean integrated squared error for the optimally tuned kernel density estimate is $n^{4/(K+1)}$ \citep{wand1995kernel}, we see similar non-parametric (polynomial decay but different exponent)  deterioration type in the convergence rate of our estimator as $K$ increases. Adding additional structural constraints discussed in Section 2.3 can greatly improve this convergence rate but the resultant estimator might be highly sub-optimal under misspecification, i.e., when the structural constraints introduced in the model is not true for the data generation process.  
		We provide further discussions on the convergence rate of our proposed estimator, as well as its implications for transfer learning, in Section~\ref{sec.r_n}.}
\end{remark}
Next we sketch the outline of and main ideas behind the proof of Theorem~\ref{thm:empirical_bound}; detailed arguments are provided in the supplement. Consider 
\begin{equation*}
{\Delta}_{\lambda,n}:=\mathbb{E} \left\{d_{\lambda}(\bm{h}_{\lambda,n},\mathfrak{h}_f)\right\}
=\mathbb{E}_{\bm{X}_n} \left[K_\lambda(\bm{x}_1,\bm{x}_n) \left\{\hat{\bm{h}}_{\lambda,n}(1) - {\mathfrak{h}}_{f} (\bm{x}_1)\right\}\, \left\{\hat{\bm{h}}_{\lambda,n}(n) - {\mathfrak{h}}_{f} (\bm{x}_n)\right\}\right],
\end{equation*} 
where the expectation is taken over $\bm{X}_n=\{\bm{x}_1,\ldots,\bm{x}_n\}$ and $\bm x_i$ are i.i.d. samples  from $f$. From \eqref{eq:k-norm} it follows that ${\Delta}_{\lambda,n} =O(n^{-1})$. 
For $\lambda \to 0$,  $K_\lambda(\bm{x}_1,\bm{x}_n)$ is negligible {only when} $||\bm{x}_1 -\bm{x}_n||_2$ is small. Thus for studying the asymptotic behavior of ${\Delta}_{\lambda,n}$, we shall restrict ourselves on the event where $||\bm{x}_1 -\bm{x}_n||_2$ is small. Conditional on this event, we show that ${\Delta}_{\lambda,n}$ can be well approximated by $\kappa_{\lambda,n} \bar{\Delta}_{\lambda,n-1}$, where $\bar{\Delta}_{\lambda,n-1} = \mathbb{E}_{\bm{X}_{n-1}} \{ (\hat{\bm{h}}_{\lambda,n}(1) - {\mathfrak{h}}_{f} (\bm{x}_1))^2 f(\bm{x}_1)\}$ and the expectation is taken over $\bm{X}_{n-1}=\{\bm{x}_1,\ldots,\bm{x}_{n-1}\}$. To heuristically understand the genesis of $\bar{\Delta}_{\lambda,n-1}$, substitute $\bm{x}_1+\bm{\epsilon}$ in place of $\bm{x}_n$ in the expression:
$${\Delta}_{\lambda,n}=\int K_\lambda(\bm{x}_1,\bm{x}_n) (\hat{\bm{h}}_{\lambda,n}(1) - {\mathfrak{h}}_{f} (\bm{x}_1))\, \big(\hat{\bm{h}}_{\lambda,n}(n) - {\mathfrak{h}}_{f} (\bm{x}_n)\big) f(\bm{x}_1)\ldots f(\bm{x}_n) \, \mathrm{d}\bm{x}_1\ldots \mathrm{d}\bm{x}_n$$
and let $|\bm{\epsilon}| \to 0$. As $\lambda \to 0$, the contributions from the kernel weight $K_\lambda$ can be separated out of the expression and subsequently accounted by constants $\kappa_{\lambda,n}$. Meanwhile the remaining terms produce $\bar{\Delta}_{\lambda,n-1}$. The rate at which $|\bm{\epsilon}| \to 0$ needs to be appropriately tuned with $\lambda$ to get the optimal rate of convergence; a rigorous probability argument is provided in the supplement. We shall see that the intermediate quantity $\bar{\Delta}_{\lambda,n-1}$, which links the $L_p$ and RKHS norms, can be explicitly characterized. The rate of convergences will be established by sandwiching $\bar{\Delta}_{\lambda,n-1}$ with functionals involving $L_1$ and $L_2$ norms.  

Finally we present a result investigating the performance of the NIT estimator under the mean squared loss. Using sub-exponential tail bounds, the $\ell_2$ loss of score functions can be obtained by extending the results on $\ell_1$ loss. The difference in the mean squared losses between the oracle and data-driven NIT estimators can be subsequently characterized.  Lemma~\ref{eq:mse_bound} below shows that this difference is asymptotically negligible. 
\begin{lemma}\label{eq:mse_bound}
	For any unknown prior $\Pi$ satisfying Assumption 1 and $\lambda \asymp n^{-1/(K+2)}$, 
	\begin{align}\label{eq:rate-3}
	\mathcal{L}_n^2 (\hat{\bm{\delta}}_{\lambda}^{\sf IT}, \bm{\theta}) -   \vphantom{\hat{\bm{\delta}}^{\text{neb}}} \mathcal{L}_n^2 (\bm{\delta}^{\pi}, \bm{\theta})  =   o_p(r_n^{-1}) \text{ as } n \to \infty.
	\end{align}	
\end{lemma}
Combining (\ref{eq:rate-2}) and (\ref{eq:rate-3}), we have established the asymptotic optimality of the data-driven NIT procedure by showing that it achieves the risk performance of the oracle rule asymptotically when $n\rightarrow\infty$; this theory is corroborated by the numerical studies in Section \ref{simu.sec}. 
\subsection{Benefits and caveats in exploiting auxiliary data}\label{sec.r_n}
The amount of efficiency gain of the data-driven NIT estimator depends on two factors: (a) the usefulness of the side information and (b) the precision of the approximation to the oracle. Intuitively when the dimension of the side information increases, the former increases whereas the latter deteriorates. 

Consider the Tweedie estimator ${y}_i+\sigma^2 \nabla \log f_1(y_i)$ that only uses the marginal density $f_1$ of $Y$ and no auxiliary information. 
The Fisher information based on the marginal $f_1$ and the conditional density $f(y|\bm{s})$ are 
$$
I_Y = \int \left\{ \frac{f_1'(y)}{f_1(y)} \right\}^2 f_1(y)\,dy \text{ and } I_{Y|\bm{S}} = \int \left\{ \frac{\nabla_y f(y|\bm{s})}{f(y|\bm{s})} \right\}^2 f(y, \bm{s})\,dy\,d\bm{s}~.
$$
The following proposition, which follows from  \cite{Bro71} (for completeness a proof is provided in the supplement), shows that, under the oracle setting, utilizing side information is always beneficial, and the efficiency gain becomes larger when more columns of auxiliary data are incorporated into the estimator. 
\begin{proposition}\label{theorem:gain}
	Consider hierarchical model (\ref{model1})--(\ref{commonInf}). Let $\bm{\delta}^\pi(\bm{y})$ and $\bm{\delta}^{\pi}(\bm{y}, \bm{S})$ respectively denote the oracle estimator with only $\bm y$ and the oracle estimator with both $\bm y$ and $\bm S$. The efficiency gain due to usage of auxiliary information is
	$$
	B_n\left\{\bm{\delta}^\pi(\bm{y})\right\} - B_n\left\{\bm{\delta}^{\pi}(\bm{y}, \bm{S})\right\} = \sigma_y^4 \big( I_{(Y|\bm{S})} - I_Y \big) \geq 0~.
	$$
	The above equality is attained if and only if the primary variable  is independent of all auxiliary variables. 
\end{proposition}
Theorems \ref{thm:empirical_bound} and Lemma \ref{eq:mse_bound} demonstrate that as the dimension $K$ increases, the rate of convergence $r_n$ decreases. This means that while adding more columns of auxiliary data (even if they are non-informative) theoretically never leads to a loss, there is still a tradeoff under our estimation framework. Specifically, the increase of $K$ can widen the gap between the oracle and data-driven rules and potentially offset the benefits of including additional side information. To better understand this tradeoff, we present a numerical example that highlights two key aspects of the phenomenon.

\begin{figure}[!h]
	\centering
	\includegraphics[width=0.7\textwidth]{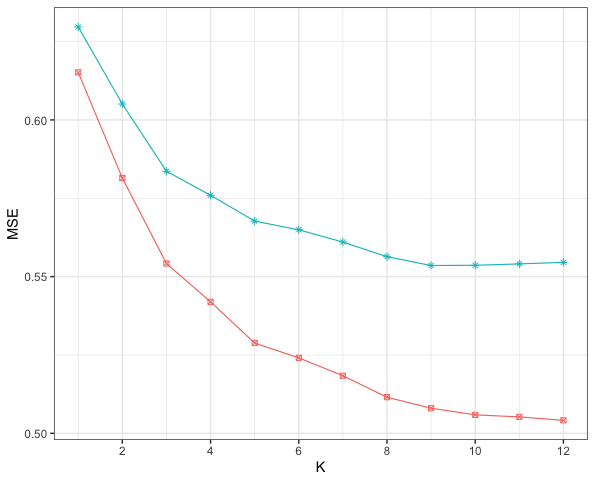}
	\caption{Mean squared error (MSE) of our proposed method (in sky blue) is plotted along with the oracle risk (in magenta) as the number of auxiliary variable $(K)$ increases. The MSE of the oracle procedure always decreases but the MSE of the data-driven NIT procedure stops decreasing as $K\geq 9$.}
	\label{fig:benefit_caveats}
\end{figure}

Consider the hierarchical model \eqref{model1}--\eqref{commonInf}. We draw the latent vector $\bm{\xi}$  from  a two-point mixture model, with equal probabilities on two atoms $0$ and $2$, \ie $\xi_i \sim 0.5\delta_{\{0\}} + 0.5\delta_{\{2\}}$. The mean vectors are simulated as $\theta_i = \xi_i + \eta_{y, i}$ and $\mu_{k, i} = \xi_i + \eta_{k, i}$, $1\le k \le K$ with $ \eta_{y, i}, \eta_{k, i} \overset{\iid}{\sim} \cN(0, 1)$. Finally we generate $Y_i \sim \cN(\theta_i, 1)$ and $S_{k, i} \sim \cN(\mu_{k,i}, 1)$, $1\le k \le K$. We vary $K$ from 1 to 12 and compare the oracle and data-driven NIT procedures in Figure \ref{fig:benefit_caveats}. We can see that the increase of $K$ has two effects: (a) the MSE of the oracle NIT procedure decreases steadily, while (b) the gap between the oracle and data-driven NIT procedures increases quickly. The combined effect initially leads to a rapid decrease in the MSE of the data-driven NIT procedure, but the decline slackens as $K \geq 5$. 

\red{In light of the above discussion, it follows that when we have a large number of auxiliary variables, it may be beneficial to conduct compress the auxiliary data to lower dimensions before applying the NIT estimator. Another remedy can be to impose structural constraints such as monotonicity or lower-dimensional functional relationship between the primary and the auxiliary data akin to \citep{ignatiadis2019empirical}.} 
\subsection{Convergence rates for heavy-tail densities}\label{sec.cr.fattail}
In this section, we expand upon the results presented in Section \ref{sec.cr.subgaussian} to encompass a broader range of prior distributions. We still assume that the true $(K+1)$-dimensional joint density $f$ as well as its score function $\mathfrak{h}_f$ are Lipschitz continuous. Akin to conditions in Theorem 5.1 of \citet{Xieetal12} we further assume that the density has $(2+\delta)$ moment bounded for some $\delta>0$. 

\smallskip
\noindent \textbf{Assumption 2:} \red{For some $\delta >0$,  the $K + 1$ dimensional joint density $f$ has bounded $2+\delta$ moment, i.e.,  $\mathbb{E}_{\bm x \sim f} \norm{\bm x}^{2+\delta} < \infty$.}
\smallskip

\noindent The next theorem shows that, for suitably chosen bandwidth, the data-driven NIT estimator is asymptotically close to the oracle estimator and the difference in their losses also converges to $0$ under any prior satisfying Assumption 2. The rate of convergence is slower than that of Theorem~\ref{thm:empirical_bound}, which is mainly due to the larger terms needed to bound heavier tails. Similar to Theorem~\ref{thm:empirical_bound}, the rate decreases with the increase of $K$.   
\begin{theorem}\label{thm:empirical_bound_heavy_tail}
	Under Assumption 2, with $\lambda \asymp n^{-1/(K+2)}$ and $$r_n=n^{{\delta}{(K+2)^{-1}(K+3+2\delta)^{-1}}}(\log n)^{-K-3},$$ we have
	$$
	r_n \cdot \bigg(\frac{1}{n}\sum_{i=1}^n \vert \hat{\bm{h}}_{\lambda,n}(i) - \mathfrak{h}_{f}(\bm{x}_i) \vert \bigg)  \to 0 ~ \text{ in } L_2 \text{ as } n \to \infty.
	$$
	Additionally, we have
	$\mathcal{L}_n^2 (\hat{\bm{\delta}}^{\sf IT}_{\lambda}, \bm{\theta}) -   \vphantom{\hat{\bm{\delta}}^{\text{neb}}} \mathcal{L}_n^2 (\bm{\delta}^{\pi}, \bm{\theta})  =   o_p(r_n^{-1}) \text{ as } n \to \infty.$
\end{theorem}
\red{The rate $r_n$ above converges to the rate in Theorem~\ref{thm:empirical_bound} as $\delta \to \infty$, which heuristically translates to the existence of all possible moments. Also, theorems~\ref{thm:empirical_bound} and \ref{thm:empirical_bound_heavy_tail} are based on the same value of the bandwidth $\lambda$.}
\subsection{Consistency of the MCV criterion}\label{sec.ddbandwidth}
In sections \ref{sec.cr.subgaussian} and \ref{sec.cr.fattail}, we have established asymptotic risk properties of our proposed method as bandwidth $\lambda \to 0$. For finite sample sizes, it is important to select the ``best'' bandwidth based on a data-driven criterion as provided in Section \ref{sec:bandwidth}. The following proposition establishes the consistency of the validation loss to the true loss, justifying the effectiveness of the bandwidth selection rule. 
\begin{proposition}\label{prop.bandwidthconsist}
	For any  fixed $\lambda>0$ and $n$,  we have
	$$
	\lim_{\alpha \to 0}  \mathbb{E} \Big\{ \hat{L}_n(\lambda, \alpha)  - \mathcal{L}_n^2 (\hat{\bm{\delta}}^{\sf IT}_\lambda, \bm{\theta}) \Big\} = 0~.
	$$
	provided that there is a unique solution to \eqref{optiprob} for $\alpha=0$. 
\end{proposition}
\section{Simulation}\label{simu.sec}
We consider three different settings where the structural information is encoded in (a) one \emph{given} auxiliary sequence that shares structural information with the primary sequence through a common latent vector (Section \ref{sec.simul.onesample}); (b) one auxiliary sequence carefully \emph{constructed  within} the same data to capture the sparsity structure of the primary sequence (Section S8.1 of the Supplement); (c) \emph{multiple} auxiliary sequences that share a common structure with the primary sequence (Section \ref{sec.simul.mutiplesample}).  

The following methods are considered in the comparison: (a) James-Stein estimator (JS); (b) the empirical Bayes Tweedie (EBT) estimator implemented using kernel smoothing as described in \cite{BroGre09}; (c) the NPMLE method by \citealp{KoeMiz14}, implemented by the R-package \texttt{REBayes} in \cite{RogJia17}; (d) the empirical Bayes with cross-fitting (EBCF) method by \cite{IgnWag19}; (e) the oracle NIT procedure \eqref{oracle} with known $f(y|\bm s)$ (NIT.OR); (f) the data-driven NIT procedure \eqref{eq:class} by solving the convex program (NIT.DD). 
The last three methods, which utilize auxiliary data, are expected to outperform the first three methods when the side information is informative. The MSE of NIT.OR is provided as the optimal benchmark for assessing the efficiency of various methods.

To implement NIT.DD, we employ the generalized Mahalanobis distance, as discussed in Section \ref{sec:bandwidth}, to compute the RBF kernel with bandwidth $\lambda$. To select an optimal $\lambda$, we solve optimization problems \ref{optiprob} across a range of $\lambda$ values and then compute the corresponding modified cross-validation (MCV) loss. The data-driven bandwidth is chosen as the value of $\lambda$ that minimizes the validation loss \eqref{valid-loss}.  

\subsection{Simulation 1: integrative estimation with one auxiliary sequence}\label{sec.simul.onesample}
\begin{figure}[!t]
	\begin{subfigure}{0.5\textwidth}
		\centering
		\includegraphics[width=0.9\textwidth]{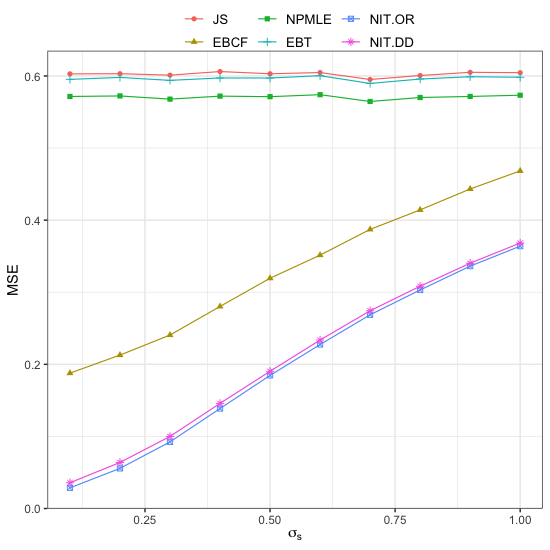}
		\caption{Setting 1}
	\end{subfigure}
	\begin{subfigure}{0.5\textwidth}
		\centering
		\includegraphics[width=0.9\textwidth]{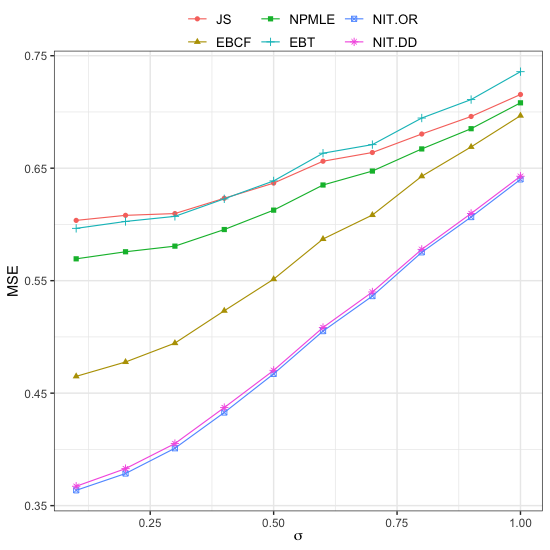}
		\caption{Setting 2}
	\end{subfigure}\\[1ex]
	\begin{subfigure}{0.5\textwidth}
		\centering
		\includegraphics[width=0.9\textwidth]{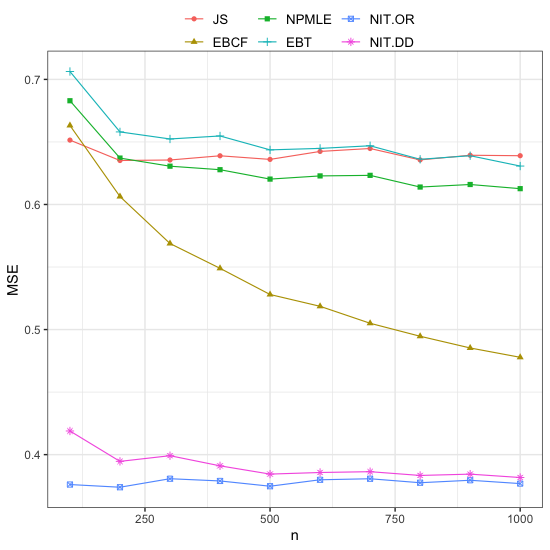}
		\caption{Setting 3}
	\end{subfigure}
	\begin{subfigure}{0.5\textwidth}
		\centering
		\includegraphics[width=0.9\textwidth]{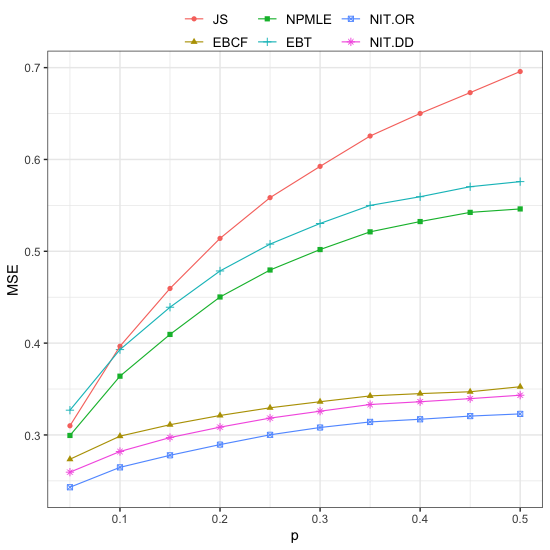}
		\caption{Setting 4}
	\end{subfigure}	
	\caption{Simulation results for one given auxiliary sequence.}\label{fig:one_sample}
\end{figure}

Let $\bm \xi=(\xi_i: 1\leq i\leq n)$ be a latent vector obeying a two-point normal mixture:
\begin{align*}
\xi_i \sim 0.5 \cN(0,1 ) + 0.5 \cN(1,1 ). 
\end{align*}
The primary data $\bm Y=(Y_i: 1\leq i\leq n)$ in the target domain are simulated according to the following hierarchical model:
$
\theta_i \sim \cN(\xi_i, \sigma^2), \quad Y_i \sim \cN(\theta_i, 1).
$
By contrast, the auxiliary data $\bm S=(S_i: 1\leq i\leq n)$ obeys
$
\zeta_i \sim  \cN(\xi_i, \sigma^2), \quad S_i \sim \cN(\zeta_i, \sigma^2_s).
$
This data generating mechanism is a special case of the hierarchical model \eqref{commonInf} where both the primary parameter $\theta_i$ and auxiliary parameter $\zeta_i$ are related to a common latent variable $\xi_i$, with $\sigma$ controlling the amount of common information shared by $\theta_i$ and $\zeta_i$. We further use $\sigma_s$ to reflect the noise level when collecting data in the source domain. The auxiliary sequence $\bm S$ becomes more useful when both $\sigma$ and $\sigma_s$ decrease. 
We consider the following settings to investigate the impact of $\sigma$, $\sigma_s$ and sample size $n$ on the performance of different methods.
\begin{description}
	\item Setting 1: we fix $n=1000$ and $\sigma\equiv 0.1$, then vary $\sigma_s$ from $0.1$ to $1$.
	\item Setting 2: we fix $n=1000$ and $\sigma_s\equiv 1$, then vary $\sigma$ from $0.1$ to $1$. 
	\item Setting 3: we fix $\sigma_s \equiv 0.5$ and $\sigma \equiv 0.5$, then vary $n$ from $100$ to $1000$.  
\end{description}
Finally we consider a setup where the auxiliary sequence is a binary vector. In the implementation of NIT.DD for categorical variables, we use indicator function to compute the pairwise distance between categorical variables. Precisely, assume that $s_i$ and $s_j$ are two categorical variables, then the distance $d(s_i, s_j) = \mathbf{1}(s_i = s_j)$.
\begin{description}
	\item Setting 4: Let $\bm \xi=(\xi_i: 1\leq i\leq n)$ be a latent vector obeying a Bernoulli distribution $\xi_i\sim {\rm Bernoulli}(p)$. The primary sequence in the target domain is generated according to a hierarchical model:
	$
	\theta_i \sim \cN(2\xi_i, 0.25), \quad y_i  \sim \cN(\theta_i, 1).
	$
	The auxiliary vector is a noisy version of the latent vector:
	$s_i \sim (1-\xi_i){\rm Bernoulli}(0.05) + \xi_i {\rm Bernoulli}(0.9)$. We fix $n=1000$ and vary $p$ from $0.05$ to $0.5$.
\end{description}				
We apply different methods to simulated data generated by the models described above and calculate the MSEs over 100 replications. Figure \ref{fig:one_sample} presents the simulation results for Settings 1-4, from which we observe several important patterns. First, the integrative methods (NIT.DD, EBCF) outperform univariate methods (JS, NPMLE, EBT) that do not incorporate auxiliary information in most settings. {Moreover,} NIT.DD consistently outperforms EBCF, with substantial efficiency gains observed in many cases. {This is not unexpected since under the data generating scheme of Simulation 1, the conditional distribution of $\theta_i$ given $S_i$ under Settings 1-4 is not necessarily Gaussian and this represents a deviation from the hierarchical model of \cite{ignatiadis2019empirical} upon which EBCF relies.} Second, the efficiency gain of the integrative methods decreases as $\sigma$ and $\sigma_s$ increase (i.e., when the auxiliary data become less informative or more noisy), as indicated by Settings 1-2. Third, as shown in Setting 3, sample size has a significant impact on integrative empirical Bayes estimation, with larger sample sizes being essential for effectively integrating side information. EBCF may under-perform univariate methods when $n$ is small. Fourth, the gap between NIT.OR and NIT.DD narrows as $n$ increases. Finally, Setting 4 demonstrates that side information can be highly informative even when the types of primary and auxiliary data differs.   
\subsection{Simulation 3: integrative estimation with multiple auxiliary sequences}\label{sec.simul.mutiplesample}
This section considers a setup where auxiliary data are collected from multiple source domains. Denote $\bm Y$ the primary sequence and $\bm S^j$, $1\leq j\leq 4$, the auxiliary sequences. In our simulation, we assume that the primary vector $\bm\theta_Y=\mathbb E(\bm Y)$ share some common information with auxiliary vectors $\bm\theta_S^j=\mathbb E(\bm S^j)$, $1\le j \le 4$ through a  latent vector $\bm\eta$, which obeys a mixture model with two point masses at 0 and 2 respectively:    
$$
\eta_i \sim 0.5\delta_{\{0\}} + 0.5\delta_{\{2\}}, \quad 1\leq i\leq n.
$$ 
There can be various ways to incorporate auxiliary data from multiple sources. We consider, in addition to NIT.DD that utilizes all sequences, an alternative strategy that involves firstly constructing a new auxiliary sequence $\bar{\bm{S}} = \frac14 (\bm{S}^1 + \bm{S}^2 + \bm{S}^3 + \bm{S}^4)$ to reduce the dimension and secondly applying NIT.DD to the pair $(\bm Y, \bar{\bm S})$; this strategy is denoted by NIT1.DD. Intuitively, if all auxiliary sequences share identical side information, then data reduction via $\bar{\bm{S}}$ is lossless.  However, if the auxiliary data are collected from heterogeneous sources with different structures and measurement units, then NIT1.DD may distort the side information and lead to substantial efficiency loss.  

To illustrate the benefits and caveats of different data combination strategies, we first consider the scenario where all sequences share a common structure via the same latent vector (Settings 1-2). Then we  turn to the scenario where the auxiliary sequences share information with the primary data in distinct ways (Settings 3-4). In all simulations below we use $n=1000$ and $100$ replications. 
\begin{description}	\item Setting 1: The primary and auxiliary data are generated from the following models: 
	\begin{equation}\label{simu-multi}
	Y_i=\theta^Y_i+\epsilon_i^Y, \quad S_i^j=\theta_{i}^{j}+\epsilon_{i}^j,
	\end{equation}
	where $\theta^Y_i\sim \cN (\eta_i, \sigma^2)$, $\theta^j_{i} \sim \cN (\eta_i, \sigma^2)$,  $1\le j \le 4$, $\epsilon^Y_i \sim \cN(0, 1)$ and $\epsilon^j_i \sim \cN(0, \sigma_s^2)$, $1\leq i\leq n$. We fix $\sigma = 0.5$ and vary $\sigma_s$ from $0.1$ to $1$.
	\item Setting 2: the data are generated using the same models as in Setting 1 except that we fix $\sigma_s = 0.5$ and vary $\sigma$ from $0.1$ to $1$.
	\item Setting 3: We generate $\bm Y$ and $\bm S^j$ using model \eqref{simu-multi}. However, we now allow $\bm\theta^j$ to have different structures across $j$. Specifically, let
	$
	\bm{\eta}^1[1:500] = \bm{\eta}[1:500],~\bm{\eta}^1[501:n] = 0,~ 
	\bm{\eta}^2[1:500] = 0$ and $\bm{\eta}^2[501:n] = \bm{\eta}[501:n]$. The following construction implies that only the first two sequences are informative in inference: 
	$$
	\theta^Y_{i} \sim \cN (\eta^1_i, \sigma^2); \quad \theta^j_{i} \sim \cN (\eta^1_i, \sigma^2), j = 1, 2;  \quad \theta^j_{i} \sim \cN (\eta^2_i, \sigma^2), j = 3, 4.
	$$ 
	We fix $\sigma = 0.5$ and vary $\sigma_s$ from $0.1$ to $1$.  
	\item Setting 4: the data are generated using the same models as in Setting 3 except that we fix $\sigma_s = 0.5$ and vary $\sigma$ from $0.1$ to $1$.
\end{description}
\begin{figure}[!h]
	\label{Simul3}
	\begin{subfigure}{0.5\textwidth}
		\centering
		\includegraphics[width=0.9\textwidth]{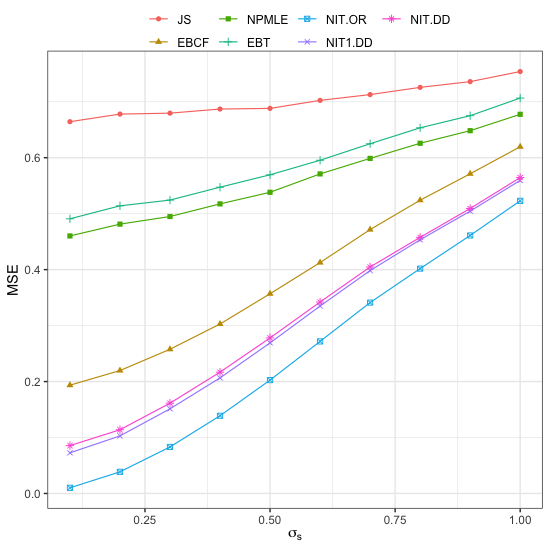}
		\caption{Setting 1}
		\label{fig:sub1}
	\end{subfigure}%
	\begin{subfigure}{0.5\textwidth}
		\centering
		\includegraphics[width=0.9\textwidth]{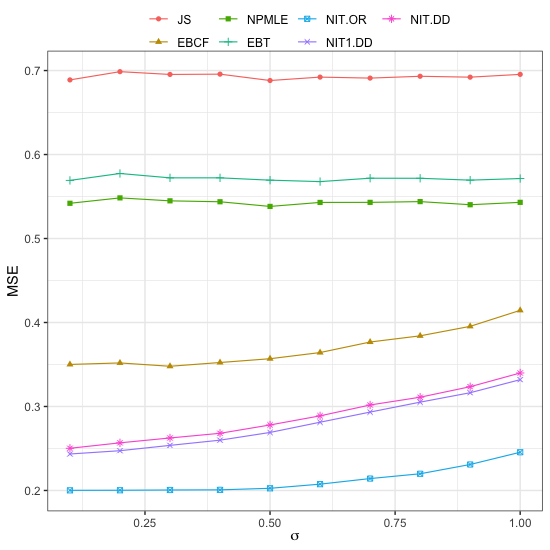}
		\caption{Setting 2}
		\label{fig:sub2}
	\end{subfigure}\\[1ex]
	\begin{subfigure}{0.5\textwidth}
		\centering
		\includegraphics[width=0.9\textwidth]{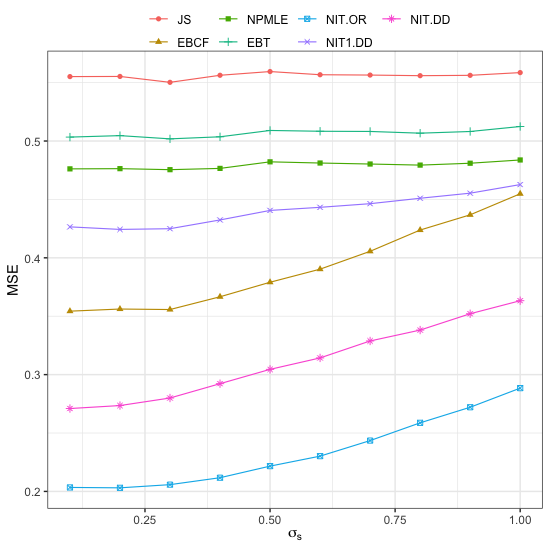}
		\caption{Setting 3}
		\label{fig:sub3}
	\end{subfigure}
	\begin{subfigure}{0.5\textwidth}
		\centering
		\includegraphics[width=0.9\textwidth]{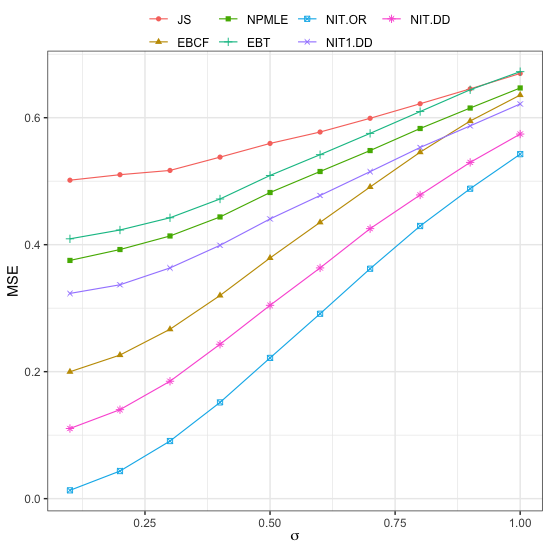}
		\caption{Setting 4}
		\label{fig:sub4}
	\end{subfigure}
	\caption{Integrative estimation with multiple auxiliary sequences. }
	\label{multi_sample}
\end{figure}		
We apply different methods to simulated data and summarize the results in Figure \ref{multi_sample}. Our observations are as follows. First, the integrative methods (NIT.DD, NIT.OR, EBCF, NIT1.DD) outperform the univariate methods (JS, NPMLE, EBT), with the efficiency gain being more pronounced when $\sigma$ and $\sigma_s$ are small. Second, NIT.DD dominates EBCF, and the gap between the performances of NIT.OR and NIT.DD widens with higher-dimensional estimation problems involving multiple auxiliary sequences. Third, in Settings 1-2, NIT1.DD is more efficient than NIT.DD as there is no loss in data reduction and fewer sequences are utilized in estimation. Finally, in Settings 3-4, the average $\bar{\bm S}$ does not provide an effective way to combine the information in auxiliary data. Improper data reduction leads to substantial information loss, such that NIT1.DD still outperforms univariate methods but is much worse than EBCF and NIT.DD. Overall, our simulation results suggest that reducing the dimension of auxiliary data can be potentially beneficial, but there can be significant information loss if the data reduction step is carried out improperly. It would be of interest to develop principled methods for data reduction to extract structural information from a large number of auxiliary sequences.

{In Section S8.1 of the supplement we present an additional simulation study involving integrative estimation in two-sample inference of sparse
	means.}
\section{Application: Integrative Nonparametric estimation of Gene Expressions} \label{app.sec}	
We consider the data set in {\cite{sen2018distinctive}} that measures gene expression levels from cells that are without interferon alpha (INFA) protein and have been infected with varicella-zoster virus (VZV). VZV is known to cause chickenpox and shingles in humans {\citep{zerboni2014molecular}}.
INFA helps in host defense against VZV but is often regulated in the presence of virus.  Thus, it is important to estimate the gene expressions in infected cells without INFA. Let $\bm{\theta}$ be the true unknown vector of mean gene expression values that need to be estimated.  Further details about the dataset is provided in Section S8.2 of the Supplement. 

\begin{figure}[!h]
	\includegraphics[width=1\textwidth]{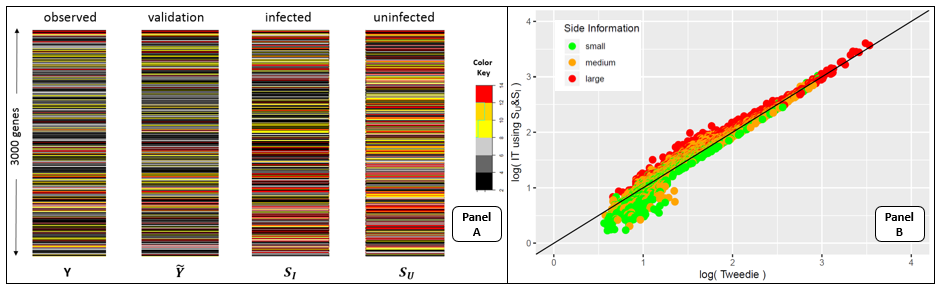}
	\caption{Panel A: Heatmaps of the gene expression datasets showing the four expression vectors corresponding to the observed, validation and auxiliary sequences. Panel B: scatterplot of the effect size estimates of gene expressions based on Tweedie and NIT (using both $S_{\sf U}$ and $S_{\sf I}$). Magnitude of the auxiliary variables used in the NIT estimate is reflected by  different colors.  
	}
	\label{fig:viro-1}
\end{figure}

The data had  gene expression measurements from two independent experiments studying VZV infected cells without INFA. We use one vector, denoted $\bm{Y}$, to construct the estimates and the other, denoted $\tilde{\bm{Y}}$, for validation. To estimate $\bm{\theta}$, alongside the primary data $\bm{Y}$, we also consider auxiliary information: $\bm{S}_{\sf U}$ which are corresponding gene expression values from uninfected cells, and 
Figure~\ref{fig:viro-1} shows the heatmap of the primary, the auxiliary and the validation sequences. We implemented the following estimators (a) the modified James-Stein (JS) following \cite{Xieetal12},
(b)  Non-parametric Tweedie  estimator without auxiliary information, (c) Empirical Bayes with cross-fitting (EBCF) by \cite{IgnWag19} and the Non-parametric Integrated Tweedie (NIT) with auxiliary information: (d) with $\bm{S}_{\sf U}$ only, (e) with $\bm{S}_{\sf I}$ only (f) using both auxiliary sequences. The mean square prediction errors of the above estimates were computed with respect to the validation vector $\tilde{\bm{Y}}$.  

Table~\ref{tab:viro1} reports the percentage gain acheived over the naive unshrunken estimator that uses $\bm{Y}$ to estimate $\bm{\theta}$. It shows that non-parametric shrinkage produces an additional 0.6\% gain over parametric JS and using auxiliary information via NIT yields a further 5.2\% gain. In particular, NIT method outperforms EBCF, which also leverage side information from both $S_{\sf U}$ and $S_{\sf I}$, by 1.7\% gain. Panel B of Figure~\ref{fig:viro-1} shows the differences between the Tweedie and NIT estimates. The differences are more pronounced in the left tails where Tweedie estimator is seen to overestimate the levels compared to NIT. The JS and NIT effective size estimates disagree by more than 50\%  at $28$ genes (which are listed in the top panel of Figure 2 in the Supplement). These genes impact $35$ biological processes and $12$ molecular functions in human cells (see bottom two panels of Figure 2 in the Supplement); this implies that important inferential gains can be made by using auxiliary information via our proposed NIT estimator.    
\begin{table}[!h]
	\centering
	\caption{\% gain in prediction errors by different estimators over the naive unshrunken estimator of gene expressions of INFA regulated infected cells.}
	\scalebox{0.8}{
		\begin{tabular}{*7c}
			\hline
			Methods &  James-Stein & Tweedie & EBCF using $\bm{S}_{\sf U}$ \& $\bm{S}_{\sf I}$ & NIT using $\bm{S}_{\sf U}$  & NIT using $\bm{S}_{\sf I}$ & NIT using $\bm{S}_{\sf U}$ \& $\bm{S}_{\sf I}$\\   
			\hline
			\% Gain & 3.5 & 4.1 & 7.6 & 6.9  & 7.5 & 9.3 \\ 
			\hline
			MSE & 2.014 & 2.001 & 1.927 & 1.930  & 1.951 & 1.895\\
			\hline
		\end{tabular}
		\label{tab:viro1}
	}
\end{table}

{In Section S8.3 of the supplement we present an additional real data application that involves leveraging auxiliary information to predict monthly sales of common grocery items across stores.}
\section{Discussion}
The NIT procedure introduces a powerful framework for leveraging useful structural knowledge from auxiliary data to facilitate the estimation of a high-dimensional parameter. It builds upon classical empirical Bayes ideas and significantly expands upon them, allowing for the handling of multivariate auxiliary data. The framework is highly versatile as it does not impose any distributional assumptions on auxiliary data, which can be categorical, numerical, or of mixed types. 

{When the noise variance $\sigma^2$ is known in Equation \eqref{model1}, 
}our theoretical analysis quantifies the reduction in estimation errors and deterioration in learning rates as the dimension of $\bm S$ increases. This suggests that when faced with a large number of variables as potential choices for auxiliary data, it may be beneficial to conduct data reduction prior to applying the NIT estimator. However, our simulation results in Section \ref{sec.simul.mutiplesample} demonstrate that improper data reduction can lead to significant information loss. This highlights the need for future research in three directions: {(a) extending the theoretical analysis considered here with a consistent estimator of $\sigma^2$,} (b) investigating the trade-off between achievable error limits of the oracle rule and the decreased convergence rate of the data-driven rule as $K$ increases, and (c) developing principled structure-preserving dimension reduction methods under the integrative estimation framework to extract useful structural information from a large number of auxiliary sequences.
\newpage
\appendix
\begin{center}
	\Large \bf{Supplement to: Empirical Bayes Estimation with Side Information: A Nonparametric
		Integrative Tweedie Approach}
\end{center}
In Sections S.1-S.7 of this supplement we present the proofs of all results stated in the main paper. The proofs are presented in the order the results appear in the main paper. We also provide an additional numerical experiment, details regarding the real data example of Section 5 and an additional real data example in Section S.8. 
\setcounter{section}{0}
\setcounter{equation}{0}
\def\theequation{S\arabic{section}.\arabic{equation}}
\def\thesection{S\arabic{section}}
\fontsize{12}{14pt plus.8pt minus .6pt}\selectfont
\vspace{10pt}
\input{append-proof}

\input{append-real}

\clearpage
\bibliographystyle{chicago}
\bibliography{Bibfiles,single-cell}

\end{document}

%% file: macros.tex
\newcommand{\BALD}{\begin{aligned}}
\newcommand{\EALD}{\end{aligned}}
\newcommand{\BALDS}{\begin{aligned*}}
\newcommand{\EALDS}{\end{aligned*}}
\newcommand{\BCAS}{\begin{cases}}
\newcommand{\ECAS}{\end{cases}}
\newcommand{\BEAS}{\begin{eqnarray*}}
\newcommand{\EEAS}{\end{eqnarray*}}
\newcommand{\BEQ}{\begin{equation}}
\newcommand{\EEQ}{\end{equation}}
\newcommand{\BIT}{\begin{itemize}}
\newcommand{\EIT}{\end{itemize}}
\newcommand{\BMAT}{\begin{bmatrix}}
\newcommand{\EMAT}{\end{bmatrix}}
\newcommand{\BNUM}{\begin{enumerate}}
\newcommand{\ENUM}{\end{enumerate}}

\newcommand{\ie}{{\it i.e. }}

\newcommand{\BA}{\begin{array}}
\newcommand{\EA}{\end{array}}

\DeclareMathOperator*{\argmin}{\arg\min}

\newcommand{\iid}{\emph{i.i.d. }}

%% file: append-proof.tex
\vspace{-12pt}
\section{Examples for integrative Tweedie}\label{sec:IT-examples}
\begin{itemize} 
\item \textbf{Example 1: }\rm{
Suppose $(Y_i, S_i)$ are conditionally independent given $(\mu_{i, y}, \mu_{i, s})$. We begin by considering a scenario where $S_i$ is an independent copy of $Y_i$:
$
\mu_i\sim N(\mu_0, \tau^2), \quad Y_i=\mu_i+\epsilon_i, \quad S_i=\mu_i+\epsilon_i^\prime, 
$
where $\epsilon_i\sim N(0, \sigma^2)$, and $\epsilon_i^\prime\sim N(0, \sigma^2)$. Intuitively, the optimal Bayes estimator is to use $Z_i=(Y_i+S_i)/2\sim N(\mu_i, \sigma^2/2)$ as the new data point:
\begin{equation}\label{opt-est}
 \hat\mu_{i}^{op} = \frac{\frac 1 2\sigma^2\mu_0+\tau^2 Z_i}{\frac 1 2\sigma^2+\tau^2}= \frac{\sigma^2\mu_0+\tau^2 (Y_i+S_i)}{\sigma^2+2\tau^2}.
\end{equation}
The conditional distribution of $Y_i$ given $S_i$ is 
$
Y_i|S_i \sim \mathcal N\left( \frac{\sigma^2\mu_0+\tau^2 S_i}{\tau^2+\sigma^2}, \frac{\sigma^2(2\tau^2+\sigma^2)}{\tau^2+\sigma^2} \right). 
$
It follows that 
$
l^\prime (Y_i|S_i)=\sigma^{-2}(2\tau^2+\sigma^2)^{-1}{(\tau^2 S_i+ \sigma^2\mu_0)-Y_i(\tau^2+\sigma^2)}.
$
We obtain
$
\delta_i^\pi = (\sigma^2+2\tau^2)^{-1} \{\sigma^2\mu_0+\tau^2 (Y_i+S_i)\},
$
recovering the optimal estimator \eqref{opt-est}. It is important to note that if we perturbate the model slightly, say by adding $\eta_i$ to $S_i$: 
$
 S_i=\mu_i+\eta_i+\epsilon_i^\prime, 
$
or letting $S_i=f(\mu_i)+\epsilon_i^\prime$, then averaging $Y$ and $S$ via \eqref{opt-est} may result in poor estimates. However, integrative Tweedie provides a robust data combination approach that consistently reduces the estimation risk (Proposition 2).}
\item \textbf{Example 2: }\rm{Consider a scenario where $S_i$ is a group indicator with two equal-sized groups ($S=1$ and $S=2$). The primary data follows 
$
Y_i|S_i=k \sim (1-\pi_k) N(0, 1) + \pi_k N(\mu_i,1),
$
with $\pi_1=0.01$, $\pi_2=0.4$, and $\mu_i \sim N(2, 1)$. We compare the performance of two oracle Bayes rules, namely $\delta_i^{\pi}(Y_i, S_i)$ and $\delta_i^{\pi}(Y_i)$. Calculations show that 
$
\big[ B(\delta_i^{\pi}(Y_i)) - B\{\delta_i^{\pi}(Y_i, S_i)\} \big]/ B\{\delta_i^{\pi}(Y_i)\} = 0.216,
$ indicating that incorporating $S_i$ can significantly reduce the risk. Despite the considerable difference between the distributions of $Y_i$ and $S_i$ (continuous vs. binary), integrative Tweedie remains highly effective in reducing estimation risk by leveraging the grouping structure encoded in $S_i$.}
\end{itemize}
\vspace{-10pt}
\section{Proof of Proposition~1}\label{proof:generalizedTF}
 The idea of the proof follows from \citet{Bro71}; we provide it here for completeness. 
 
{First, note that $\nabla_y \log f(y|s)=\nabla_y \log f(y,s)$ as 
$$\nabla_y \log f(y|s)=\nabla_y \{\log f(y,s)  - \log f(s)\}=\nabla_y \log f(y,s).$$ Next, from equations (1.1) and (2.2), 
$\bm S_i$ and $Y_i$ are independent given $\theta_i$, and so \(f(y|\theta, \bm{s}) = f(y|\theta)\) for all \(\theta\) and \(\bm{s}\).} Therefore, noting that $f(y, \pmb{s}) = \int f(y, \pmb{s} | \theta)dh_{\theta}(\theta)$ and $ f(y, \pmb{s} | \theta) = f(y | \theta)f(\pmb{s} | \theta)$, expand the partial derivative of $f(y, \pmb{s})$:
	\begin{align*}
		\nabla_{y}f(y, \pmb{s}) &= \sigma^{-2}\Big(\int \theta f(y |\pmb{s}, \theta)f(\pmb{s} | \theta) dh_{\theta}(\theta) - y \int f(y |\pmb{s}, \theta)f(\pmb{s} | \theta) dh_{\theta}(\theta)\Big) \\
		&= \sigma^{-2} \Big(\int \theta f(y, \pmb{s} | \theta) dh_{\theta}(\theta) - yf(y, \pmb{s})\Big)
	\end{align*}
	Then, left-multiplying by $\sigma^2$ and dividing by $f(y, \pmb{s})$ on both sides, it follows that
	\begin{align*}
		\sigma^2 \frac{\nabla_{y}f(y, \pmb{s})}{f(y, \pmb{s})} = \frac{\int \theta f(y, \pmb{s} | \theta) dh_{\theta}(\theta) }{ f(y, \pmb{s} | \theta)  }- y
	\end{align*}
	Under square error loss, the posterior mean minimizes the Bayes risk. And so, the Bayes estimator is given by
	\begin{align*}
		\E (\theta | y, \pmb{s}) = \frac{\int \theta f(y, \pmb{s} | \theta) dh_{\theta}(\theta)}{f(y, \pmb{s})} = y + \sigma^2 \frac{\nabla_{y}f(y, \pmb{s})}{f(y, \pmb{s})}~,
	\end{align*}
where, the second equality follows from the above two displays.
	
\section{Proof of Theorem~1}\label{sec.proof.thm1}
First note that the expected value of the concerned $\ell_p$ distance
$$\ell_p( \hat{\pmb{h}}_{\lambda,n},\mathfrak{h}_{f})=n^{-1}\sum_{i=1}^n \vert \hat{\pmb{h}}_{\lambda,n}(i) - \mathfrak{h}_{f}(\pmb{x}_i)|^p$$ is given by
$\Delta_{\lambda,n}^{(p)}(f)=  \mathbb{E}_{\pmb{X}} \{\ell_p( \hat{\pmb{h}}_{\lambda,n},\mathfrak{h}_{f}) \}$
where, the expected value is over $\pmb{X}_n=(\pmb{x}_1;\pmb{x}_2;\ldots;\pmb{x}_n)$ where $\pmb{x}_i$s are i.i.d. from $f$.  Thus,
$$\Delta^{(p)}_{\lambda,n}(f)= \mathbb{E}_{\,} |\hat{\pmb{h}}_{\lambda,n}(1) - \mathfrak{h}_{f}(\pmb{x}_1) |^p  = \mathbb{E} |\hat{\pmb{h}}_{\lambda}[\pmb{X}_n](\pmb{x}_1) - \mathfrak{h}_{f}(\pmb{x}_1) |^p~.$$
For notational ease, we would often keep the dependence on $f$ in $\Delta^{(p)}_{\lambda,n}(f)$ implicit. The proof involves upper and lower bounding  $\Delta^{(2)}_{\lambda,n}$ by the functionals involving $\Delta^{(1)}_{\lambda,n}$. The upper bound is provided below in \eqref{eq:a3}. The lower bound follows from \eqref{eq:a7}, whose proof is quite convoluted and is presented separately in Lemma~\ref{Technical_lemma_1}.

As the marginal density of the $\pmb{\theta}$  is the convolution with a Gaussian distribution, it follows that there exists some constant $C\geq 0$ such that $$|\mathfrak{h}_f(\pmb{x}_1)| /\norm{\pmb{x}_1}_2 \leq C \text{ for all large } ||\pmb x_1||_2.$$ 
and $|\hat{\pmb{h}}_{\lambda}[\pmb{X}_n](\pmb{x}_1) | =O(\norm{\pmb{x}_1}_2)$. With out loss of generality we include such constraints on $\pmb{h}$ in the convex program to solve (2.4) and so, $ |\hat{\pmb{h}}_{\lambda}[\pmb{X}_n](\pmb{x}_1) - \mathfrak{h}_{f}(\pmb{x}_1)|$ is also bounded by $O(\norm{\pmb{x}_1}_2)$.

Using this property of the score estimates,  we have the following bound  for all $\pmb{x}_1$ satisfying 
$\left\{\pmb x_1: \norm{\pmb{x}_1}_2 \le 2\gamma\log n \right\}:$   
\begin{align}\label{eq:a1}
	\mathbb{E} \left[  \big(\hat{\pmb{h}}_{\lambda}[\pmb{X}_n](\pmb{x}_1) - \mathfrak{h}_{f}(\pmb{x}_1)\big)^2I\left\{\norm{\pmb{x}_1}_2 \le 2\gamma\log n \right\} \right] 
	\le  2\gamma\log(n)\, \Delta^{(1)}_{\lambda,n} .
\end{align}
On the set $\left\{ \norm{\pmb{x}_1}_2 > 2\gamma\log n \right\}$, again using the aforementioned property of score estimates from (2.4) we note that
\begin{align}\label{eq:a2}
	\begin{split}
		\mathbb{E} \left[\left(\hat{\pmb{h}}_{\lambda}[\pmb{X}_n](\pmb{x}_1) - \mathfrak{h}_{f}(\pmb{x}_1)\right)^2 I\left\{\norm{\pmb{x}_1}_2 > 2\gamma\log n\right\} \right] &\lesssim  \mathbb{E} \left[ \norm{\pmb{x}_1}_2^2 I\{\norm{\pmb{x}_1}_2 > 2\gamma\log n\}\right],
	\end{split}
\end{align}
where, for any two sequences $a_n$, $b_n$, we use the notation $a_n \lesssim b_n$ to denote $a_n/b_n=O(1)$ as $n \to \infty$.  

Now, as $\pmb x_1$ satisfies assumption 1, the right hand side \eqref{eq:a2} is bounded by $O(n^{-1})$. Combining \eqref{eq:a1} and \eqref{eq:a2} we have the following upper bound on $\Delta^{(2)}_{\lambda,n}$:
\begin{align}\label{eq:a3}
	\Delta^{(2)}_{\lambda,n} \lesssim \log(n)\, \Delta^{(1)}_{\lambda,n} + n^{-1}~.
\end{align}

For the lower bound on  $\Delta^{(2)}_{\lambda,n}$ consider the following intermediate quantity which is related to the KSD norm $d_{\lambda}$ on the score functions: 
$$\bar\Delta_{\lambda,n}(f)= \mathbb{E} \big\{\big(\hat{\pmb{h}}_{\lambda}[\pmb{X}_n](\pmb{x}_1) - \mathfrak{h}_{f}(\pmb{x}_1)\big)^2 f(\pmb{x}_1)\big\}~.$$
It can be shown that
\begin{align}\label{eq:a4}
\Delta^{(1)}_{\lambda,n} \lesssim \sqrt{\{\log(n)\}^{K+1}\, \bar \Delta_{\lambda,n}}+n^{-1} \text{ as } n \to \infty. 
\end{align}

\textbf{Proof of \eqref{eq:a4}.}  Restricting $\pmb{x}_1$ on set $\left\{\pmb x_1: \norm{\pmb{x}_1}_2 \le 2\gamma\log n \right\}$ and using Cauchy-Schwarz inequality, we get 
\begin{align*}
	\begin{split}
		\mathbb{E} \left[  \big(\hat{\pmb{h}}_{\lambda}[\pmb{X}_n](\pmb{x}_1) - \mathfrak{h}_{f}(\pmb{x}_1)\big)^2I\left\{\norm{\pmb{x}_1}_2 \le 2\gamma\log n \right\} \right] 
		\le \left[ C_{K, \gamma}\, \{\log (n)\}^{K+1}\bar\Delta_{\lambda,n}(f) \right]^{\frac12}.
	\end{split}
\end{align*}
On the tail $\left\{\pmb x_1: \norm{\pmb{x}_1}_2 > 2\gamma\log n \right\}$ using the same argument as \eqref{eq:a2}, we have
\begin{align*}
		\mathbb{E} \left[\left|\hat{\pmb{h}}_{\lambda}[\pmb{X}_n](\pmb{x}_1) - \mathfrak{h}_{f}(\pmb{x}_1)\right| I\left\{\norm{\pmb{x}_1}_2 > 2\gamma\log n\right\} \right] &= O(n^{-1}).
\end{align*}
\eqref{eq:a4} follows by combining the above two displays. \\[1ex]

The following result lower bounds  $\Delta^{(2)}_{\lambda,n}$ using $\bar \Delta_{\lambda,n}$. 
\begin{lemma}\label{Technical_lemma_1}
For any $\lambda >0$, we have
	\begin{align}\label{eq:a7}
	 \bar \Delta_{\lambda,n}
	\lesssim \lambda^{-(K+1)} \mathcal{S}_{\lambda}[\hat{\pmb{h}}_{\lambda,n+1}] + \lambda^2 \log n +    \lambda  (\log n)^{K+3}  \Delta_{\lambda,n}^{(2)} ~.
	\end{align}
\end{lemma}
The proof of the above lemma is intricate and is presented at the end of this section.\\[1ex]  

Now, for the proof of Theorem~1, we combine \eqref{eq:a3}, \eqref{eq:a4} and \eqref{eq:a7}. Then, using  $\lambda \asymp n^{-\frac{1}{K+2}}$ and the fact that $\Delta^{(1)}_{\lambda,n}$ is bounded, we arrive at
\begin{align}\label{eq:a12}
	\Delta^{(1)}_{\lambda,n} \lesssim  \sqrt{\left\{\log(n)\right\}^{K+1}\, \left\{   n^{\frac{K+1}{K+2}}\mathcal{S}_{\lambda}[\hat{\pmb{h}}_{\lambda,n+1}]  + n^{-\frac{2}{K+2}}\log (n) + n^{-\frac{1}{K+2}} (\log n)^{K+4} \, \Delta^{(1)}_{\lambda,n}  \right\} }.
\end{align}
Proportion~\ref{propLoss}, which is stated and proved at the end of this proof, provides the following upper bound on  $\mathcal{S}_{\lambda}[\hat{\pmb{h}}_{\lambda,n+1}]$: 
\begin{align}\label{eq:a8}
	\mathcal{S}_{\lambda}[\hat{\pmb{h}}_{\lambda,n+1}]  \le \frac{\mathbb{E}\left\{ \mathfrak{h}_{f}(\pmb{x}_1)\right\}^2 - \mathbb{E}\left\{\hat{\pmb{h}}_{\lambda}[\pmb{X}_{n+1}](\pmb{x}_1) \right\}^2}{n}~
\end{align}
Using the similar argument as \eqref{eq:a3}, the numerator in above can be further upper bounded by $	2\gamma\,\Delta^{(1)}_{\lambda,n+1} + O(n^{-1})$. Substituting this in \eqref{eq:a12}, we arrive at an inequality only involving quantities $\Delta^{(1)}_{\lambda,n}$ and $\Delta^{(1)}_{\lambda,n+1}$. Now, noting that
 $\lambda \asymp n^{-\frac{1}{K+2}}$ and $\Delta^{(1)}_{\lambda,n}$ is bounded, it easily follows that $\Delta^{(1)}_{\lambda,n} \to 0$ as $n\to \infty$. 
 
Establishing the rate of convergence of $\Delta^{(1)}_{\lambda,n}$ needs further calculations. For that purpose consider $A_n = \max \left\{ \Delta^{(1)}_{\lambda,n},\, 2\,n^{-\frac{1}{K+2}}(\log n)^{2K+5}\right\}$. For all large $n$, the following  inequality can be derived from \eqref{eq:a12} and \eqref{eq:a8}:
\begin{align}\label{theorem_population:4}
A_n \leq C\, (\log n)^{K+1} n^{-\frac{1}{2K+4}}\sqrt{A_{n+1}},
\end{align}
where $C$ is a constant independent of $n$. 

Applying (\ref{theorem_population:4}) recursively $m$ times we have:
\begin{align*}
\begin{split}
A_n  \le \Big( C (\log n)^{K+1} n^{-\frac{1}{2K+4}} \Big)^{1+\cdots + \frac{1}{2^m}}\, A_{n+m+1}^{\frac{1}{2^{m+1}}}.
\end{split}
\end{align*}
Note that $A_n < 1$ for all large $n$. This implies that for any $m > 0$,
\begin{align*}
A_n & \le \Big( C (\log n)^{K+1} n^{-\frac{1}{2K+4}} \Big)^{1+\cdots + \frac{1}{2^m}}.
\end{align*}
Finally, let $m\to \infty$, we proved that $A_n  \le C (\log n)^{2K+2} n^{-\frac{1}{K+2}} $, which implies
\begin{align*}
 \Delta^{(1)}_{\lambda,n} \lesssim  (\log n)^{2K+2} n^{-\frac{1}{K+2}}~.
\end{align*}
This completes the proof of Theorem~1.
\subsection{Proofs of results used in the proof of Theorem~1}
\begin{proposition}\label{propLoss}
	Let $K_{\lambda}(\cdot, \cdot)$ be RBF kernel with bandwidth parameter $\lambda \in \Lambda$ and $\Lambda$ is a compact set of $\mathbb{R}^+$ bounded from zero.  Then we have
	\begin{align*}
	\mathcal{S}_{\lambda}[\hat{\pmb{h}}_{\lambda,n}]  \le \frac{\mathbb{E}\left\{ \mathfrak{h}_{f}(\pmb{x}_1)\right\}^2 - \mathbb{E}\left\{\hat{\pmb{h}}_{\lambda}[\pmb{X}_n](\pmb{x}_1) \right\}^2}{n-1}.
	\end{align*}
\end{proposition}

\noindent \textbf{Proof of Proposition \ref{propLoss}.}
	By the construction of the $\hat{\pmb{h}}_{\lambda, n}$, we have
	\begin{align}\label{eq:const_ineq}
	\widehat{\mathcal{S}}_{\lambda}[\hat{\pmb{h}}_{\lambda,n}]  \le \widehat{\mathcal{S}}_{\lambda}[\pmb{\mathfrak{h}}_f ].
	\end{align}
	Taking the expectation on the both sides of equation (\ref{eq:const_ineq}), we get
	\begin{align*}
	\frac{n^2-n}{n^2}	\mathcal{S}_{\lambda}[\hat{\pmb{h}}_{\lambda,n}]  + \frac{n}{n^2} \Big(\mathbb{E}\left\{\hat{\pmb{h}}_{\lambda}[\pmb{X}_n](\pmb{x}_1) \right\}^2 + \frac{1}{\lambda}\Big) \le \frac{n^2-n}{n^2}	\mathcal{S}_{\lambda}[\mathfrak{h}_f]    + \frac{n}{n^2} \Big(\mathbb{E}\left\{ \mathfrak{h}_{f}(\pmb{x}_1)\right\}^2 + \frac{1}{\lambda} \Big).
	\end{align*}
	Notice that $	\mathcal{S}_{\lambda}[\mathfrak{h}_f] = 0$ and then the above inequality implies
	\begin{align*}
	\mathcal{S}_{\lambda}[\hat{\pmb{h}}_{\lambda,n}]   \le \frac{\mathbb{E}\left\{ \mathfrak{h}_{f}(\pmb{x}_1)\right\}^2 - \mathbb{E}\left\{\hat{\pmb{h}}_{\lambda}[\pmb{X}_n](\pmb{x}_1) \right\}^2}{n-1},
	\end{align*}
which completes the proof. 

\subsection*{Proof of  Lemma~\ref{Technical_lemma_1}.}
First we assume there are $n+1$ i.i.d. samples, $\pmb{X}_{n+1}=(\pmb{x}_1;\pmb{x}_2;\ldots;\pmb{x}_{n+1})$ where $\pmb{x}_i$s are i.i.d. from $f$.
Note that the definition of $\mathcal{S}_{\lambda}[\hat{\pmb{h}}_{\lambda,n+1}]$ is equivalent to the following definition:
\begin{align*}
\mathcal{S}_{\lambda}[\hat{\pmb{h}}_{\lambda,n+1}] = \mathbb{E} \left[ D_{\lambda}(\pmb x_1, \pmb x_{n+1}) \right],
\end{align*}
where the KSD is given by
$$ D_{\lambda}(\pmb x_1, \pmb x_{n+1})= K_{\lambda}(\pmb x_1, \pmb x_{n+1}) \big(\hat{\pmb{h}}_{\lambda}[\pmb{X}_{n+1}](\pmb{x}_1) - \mathfrak{h}_{f}(\pmb{x}_1) \big) \big(\hat{\pmb{h}}_{\lambda}[\pmb{X}_{n+1}](\pmb{x}_{n+1}) - \mathfrak{h}_{f}(\pmb{x}_{n+1}) \big).
$$

We consider the situation when $\pmb x_{n+1}$ is in the $\epsilon$-neighboor of $\pmb x_1$. For a fixed $\epsilon > 0$, denote 
\begin{align*}
I^{(1)}_{\epsilon; \lambda} := \mathbb{E} \Big[  D_{\lambda}(\pmb x_1, \pmb x_{n+1}) I\{\norm{\pmb{x}_{n+1}-\pmb{x}_1}< \epsilon\} \Big].
\end{align*}
When $\epsilon = \lambda \log n$, we have 
\begin{align}\label{eq:a31}
I^{(1)}_{\epsilon; \lambda} \le 	\mathcal{S}_{\lambda}[\hat{\pmb{h}}_{\lambda,n+1}]  + O\left(n^{-0.5\log n} \right).
\end{align}
The proof of \eqref{eq:a31} is non-trivial. To avoid disrupting the flow of arguments here, its proof is not presented immediately but is provided at the end of this subsection.

Denote the following intermediate quantity $	I^{(2)}_{\epsilon; \lambda}$ which is close to $\bar\Delta_{\lambda,n}(f)$ as
\begin{align*}
I^{(2)}_{\epsilon; \lambda} := \E \left[K_{\lambda}(\pmb x_1, \pmb x_{n+1}) \big(\hat{\pmb{h}}_{\lambda}[\pmb{X}_{n+1}](\pmb{x}_1) - \mathfrak{h}_{f}(\pmb{x}_1) \big)^2  I\{\norm{\pmb{x}_{n+1}-\pmb{x}_1}< \epsilon\} \right].
\end{align*}
We use Cauchy Schwarz inequality and Lipschitz continuity of score function to show $	I^{(2)}_{\epsilon; \lambda}$ is bounded by a function of $	I^{(1)}_{\epsilon; \lambda}$ as
\begin{align}\label{eq:a34}
I^{(2)}_{\epsilon; \lambda}\le 	I^{(1)}_{\epsilon; \lambda}  + O(\epsilon^{K+3}).
\end{align}
The proof of \eqref{eq:a34} is quite involved and is presented afterwards. 
Finally, we establish the following bound which along with \eqref{eq:a31} and \eqref{eq:a34} complete the proof of the lemma:
\begin{align}\label{eq:a37} 
\bar \Delta_{\lambda,n} \lesssim 	\lambda^{-K-1}	I^{(2)}_{\epsilon; \lambda}  + \lambda^2 (\log n)^{K+3} + \lambda\Delta^{(2)}_{\lambda,n} \log n.
\end{align}

\textbf{Proof of \eqref{eq:a31}.} Note that the difference between $\mathcal{S}_{\lambda}[\hat{\pmb{h}}_{\lambda,n+1}]$ and 	$	I^{(1)}_{\epsilon; \lambda}$ is 
\begin{align*}
\mathbb{E} \Big[  D_{\lambda}(\pmb x_1, \pmb x_{n+1}) I\{\norm{\pmb{x}_{n+1}-\pmb{x}_1}\ge \epsilon\} \Big].
\end{align*}
If we use the Gaussian kernel $K_{\lambda}(\pmb x_1, \pmb x_{n+1}) = e^{-\frac{1}{2\lambda^2}\norm{\pmb{x}_1 - \pmb{x}_{n+1}}^2}$ and set $\epsilon = \lambda \log n$, we have  $ K_{\lambda}(\pmb x_1, \pmb x_{n+1}) I\{\norm{\pmb{x}_{n+1}-\pmb{x}_1}\ge \epsilon\}$ is always bounded by $n^{-0.5\log n}$, which implies the above difference is bounded by $\Delta^{(2)}_{\lambda,n+1} n^{-0.5\log n}$. Note that $	\Delta^{(2)}_{\lambda,n+1}$ is bounded,  \eqref{eq:a31} follows.\\[1ex]

\textbf{Proof of \eqref{eq:a34}.} Note that the score function $\mathfrak{h}_{f}$ is $L_f$-Lipschitz continuous. If we assume for small $\epsilon$, when $\norm{\pmb{x}_{n+1}-\pmb{x}_1}< \epsilon$, we have $\hat{\pmb{h}}_{\lambda}[\pmb{X}_{n+1}](\pmb{x}_{n+1}) $ is $L_{n, \epsilon}$-Lipschitz continuous as
\begin{align}\label{eq:a35}
\left|\hat{\pmb{h}}_{\lambda}[\pmb{X}_{n+1}](\pmb{x}_{n+1}) - \hat{\pmb{h}}_{\lambda}[\pmb{X}_{n+1}](\pmb{x}_1) \right| \le L_{n, \epsilon}\, \epsilon.
\end{align}
where $L_{n, \epsilon}$  satisfies that $\E L_{n, \epsilon}^2$ is bounded. Then the difference between $	I^{(2)}_{\epsilon; \lambda}$ and $	I^{(1)}_{\epsilon; \lambda}$ is bounded by 
\begin{align*}
\mathbb{E} \Big[ \epsilon\, (L_f + L_{n, \epsilon})K_{\lambda}(\pmb x_1, \pmb x_{n+1}) \left|\hat{\pmb{h}}_{\lambda}[\pmb{X}_{n+1}](\pmb{x}_1) - \mathfrak{h}_{f}(\pmb{x}_1) \right|   I\{\norm{\pmb{x}_n-\pmb{x}_1}< \epsilon\} \Big] 
\end{align*}
Apply the Cauchy-Schwarz inequality and the square of above difference can be further bounded by 
\begin{align*}
\epsilon \, \E\left[(L_f + L_{n, \epsilon})^2 I\{\norm{\pmb{x}_n-\pmb{x}_1}< \epsilon\}\right] \E \left[K^2_{\lambda}(\pmb x_1, \pmb x_{n+1}) \left|\hat{\pmb{h}}_{\lambda}[\pmb{X}_{n+1}](\pmb{x}_1) - \mathfrak{h}_{f}(\pmb{x}_1) \right|^2   I\{\norm{\pmb{x}_n-\pmb{x}_1}< \epsilon\} \right].
\end{align*}

Note that  $\E\left[(L_f + L_{n, \epsilon})^2 I\{\norm{\pmb{x}_n-\pmb{x}_1}< \epsilon\}\right] $ is bounded by $$C_f \frac{\pi^{(K+1)/2}}{\Gamma(\frac{K+1}{2}+1)}\epsilon^{K+1}\E(L_f + L_{n, \epsilon})^2,$$ where $\Gamma(x)$ is the gamma function. Notice that $K^2_{\lambda}(\pmb x_1, \pmb x_{n+1}) \le K_{\lambda}(\pmb x_1, \pmb x_{n+1})$ and then we have 
\begin{align*}
I^{(2)}_{\epsilon; \lambda}\lesssim 	I^{(1)}_{\epsilon; \lambda} + \epsilon \sqrt{\epsilon^{K+1}	\Delta^2_{\epsilon; \lambda}}.
\end{align*}
This completes the proof of \eqref{eq:a34}.\\[1ex]

\textbf{Proof of \eqref{eq:a37}.} We introduce an intermediate quantity:
\begin{align*}
I^{(3)}_{\epsilon; \lambda} = \mathbb{E} \left[ K_{\lambda}(\pmb x_1, \pmb x_{n+1}) \big(\hat{\pmb{h}}_{\lambda}[\pmb{X}_{n}](\pmb{x}_1) - \mathfrak{h}_{f}(\pmb{x}_1) \big)^2 I\{\norm{\pmb{x}_{n+1}-\pmb{x}_1}< \epsilon\} \right].
\end{align*}
Assume that when $n$ is large and $\norm{\pmb x_{n+1} - \pmb x_1} < \epsilon$, we have  $\hat{\pmb{h}}_{\lambda}[\pmb{X}_{n+1}](\pmb{x}_{n+1}) $ is $L_{n, \epsilon}$-Lipschitz continuous as:
\begin{align*}
\left|\hat{\pmb{h}}_{\lambda}[\pmb{X}_{n+1}](\pmb{x}_{n+1}) - \hat{\pmb{h}}_{\lambda}[\pmb{X}_{n}](\pmb{x}_1) \right| \le L_{n, \epsilon} \,\epsilon
\end{align*}
Combined with \eqref{eq:a35}, we get the difference between $	I^{(3)}_{\epsilon; \lambda} $ and 
$	I^{(2)}_{\epsilon; \lambda} $ is bounded by
\begin{align*}
4\epsilon^2\,\E\left[ L_{n, \epsilon}^2K_{\lambda}(\pmb x_1, \pmb x_{n+1})  I\{\norm{\pmb{x}_{n+1}-\pmb{x}_1}< \epsilon\} \right],
\end{align*}
which implies that
\begin{align}\label{eq:a41}
I^{(3)}_{\epsilon; \lambda} \lesssim 	I^{(2)}_{\epsilon; \lambda} + \epsilon^{K+3}.
\end{align}
Next we introduce another intermediate quantity 
\begin{align*}
I^{(4)}_{\epsilon; \lambda} = \E \int f(\pmb x_1) K_{\lambda}(\pmb x_1, \pmb x_{n+1}) \big(\hat{\pmb{h}}_{\lambda}[\pmb{X}_{n}](\pmb{x}_1) - \mathfrak{h}_{f}(\pmb{x}_1) \big)^2 I\{\norm{\pmb{x}_{n+1}-\pmb{x}_1}< \epsilon\} \, d\pmb x_{n+1},
\end{align*}
which is close to 	$	I^{(3)}_{\epsilon; \lambda}$. When $\epsilon = \lambda \log n$, we have 
the following term
$$\int K_{\lambda}(\pmb x_1, \pmb x_{n+1})  I\{\norm{\pmb{x}_{n+1}-\pmb{x}_1}< \epsilon\} \, d\pmb x_{n+1}$$ 
is lower bounded by 
$$
\lambda^{K+1}\int e^{-\frac12 \norm{\pmb x_{n+1}}^2} I \{ \norm{\pmb x_{n+1}} < \log n\}\, d\pmb x_{n+1}, 
$$
which can be further lower bounded by $c\,\lambda^{K+1}$ for some constant $c$ when $n$ is large. This implies 
\begin{align}\label{eq:a43}
\lambda^{K+1}\,	\bar\Delta_{\lambda,n}(f) \lesssim 	I^{(4)}_{\epsilon; \lambda}.
\end{align}
Now it is enough to show $	I^{(4)}_{\epsilon; \lambda} \lesssim 	I^{(3)}_{\epsilon; \lambda} + \lambda^{K+2} \Delta^{(2)}_{\lambda,n} \log n.$

Assume that $f$ is $L_f$-Lipschitz continuous. The difference between  $	I^{(4)}_{\epsilon; \lambda}$ and $	I^{(3)}_{\epsilon; \lambda}$ is bounded by 
\begin{align*}
L_f \epsilon\, \Delta^{(2)}_{\lambda,n} \int \left[ K_{\lambda}(\pmb x_1, \pmb x_{n+1})  I\{\norm{\pmb{x}_{n+1}-\pmb{x}_1}< \epsilon\} \right] \,d\pmb x_{n+1}.
\end{align*}
Notice that $ \int \left[ K_{\lambda}(\pmb x_1, \pmb x_{n+1})  I\{\norm{\pmb{x}_{n+1}-\pmb{x}_1}< \epsilon\} \right] \,d\pmb x_{n+1}$ is bounded by $C \,\lambda^{K+1}$ for some constant $C$. This implies that 
$$
I^{(4)}_{\epsilon; \lambda} \lesssim I^{(3)}_{\epsilon; \lambda} + \lambda^{K+2} 	\Delta^{(3)}_{\lambda,n} \log n.
$$
Combined with \eqref{eq:a41} and \eqref{eq:a43}, the result \eqref{eq:a37} follows.
\section{Proof of Lemma~1}
We follow the notions in Section~\ref{sec.proof.thm1}. The convergence rate of $\Delta^{(2)}_{\lambda,n}$ is achieved by extending the results of $\Delta^{(1)}_{\lambda,n}$ in Section~\ref{sec.proof.thm1}. Recall that \eqref{eq:a3} shows
\begin{align*}
	\Delta^{(2)}_{\lambda,n} \lesssim \log(n)\, \Delta^{(1)}_{\lambda,n} + n^{-1}~,
\end{align*}
and we have proved $\Delta^{(1)}_{\lambda,n} \lesssim  (\log n)^{2K+2} n^{-\frac{1}{K+2}}$ in Section~\ref{sec.proof.thm1}. Combining these two, we obtain the result stated in this lemma.
\section{Proof of Proposition 2}
Proposition 4.5 in \citet{johnstone2011gaussian} shows that $B_n(\pmb{\delta}^{\pi}(\pmb{y})) = \sigma^2- \sigma^4 I_Y$. Following the same arguments, we have $B_n(\pmb{\delta}^{\pi}(\pmb{y}, \pmb{S})) = \sigma^2- \sigma^4 I(p_{y|\pmb{s}})$. Then it follows
\begin{align*}
	B_n(\pmb{\delta}^{\pi}(\pmb{y}))- B_n(\pmb{\delta}^{\pi}(\pmb{y}, \pmb{S})) = \sigma^4(I_{Y|\pmb S} - I_Y)
\end{align*}
Next, we prove that $I_{Y|\pmb S} - I_Y $ is non-negative. By the definition of $I_{Y|\pmb S}$, we have the following decomposition:
\begin{align*}
	I_{Y|\pmb S} =& \iint \Big( \frac{f(y)\nabla_y f(\pmb{s}|y) + f(\pmb{s}|y) \nabla_y f(y)  }{f(y, \pmb{s})}  \Big)^2 f(y, \pmb{s})\,dy\,d\pmb{s}.
\end{align*}
Then we break the square and it follows 
\begin{align*}
	I_{Y|\pmb S} 
	= \iint \Big( \frac{\nabla_y f(\pmb{s}|y)}{f(\pmb{s}|y)}\Big)^2 f(y, \pmb{s})\,dy\,d\pmb{s} +  \iint \Big( \frac{\nabla_y f(y)  }{f(y)}  \Big)^2f(y)\,dy
	+ 2 \iint \nabla_y f(\pmb{s}|y)\nabla_y f(y)   \,dy\,d\pmb{s}.
\end{align*}
Note that the second term of right hand side is always non-negative. Then we consider the last term and exchange the integration and partial derivative, we get 
\begin{align*}	
	\iint \nabla_y f(\pmb{s}|y)\nabla_y f(y)\,dy\,d\pmb{s} &= \int \nabla_y f(y) \nabla_y\Big( \int  f(\pmb{s}|y)d\pmb{s} \Big)dy= 0
\end{align*}
It follows that $I_{Y}\ge I_{Y|\pmb S}$.
\section{Proof of Theorem~2}
The proof of this theorem follows along the similar lines of the proof for Theorem~1. Denote $\beta = \frac{1}{(K+2)(K+3+2\delta)}$. In this case we entertain the possibility that the joint density $f$ can be a heavier tailed density.  We concentrate on set $\left\{\norm{\pmb{x}_1}_2 \le  n^{\beta} \right\}$ instead of the set  $\left\{\norm{\pmb{x}_1}_2 \lesssim  \log n \right\}$ analyzed in the proof of Theorem~1.

Noting that $ \hat{\pmb{h}}_{\lambda}[\pmb{X}_n](\pmb{x}_1) $ and $ \mathfrak{h}_{f}(\pmb{x}_1)$ are both $O\left(\norm{\pmb{x}_1}\right)$, it follows that $ |\hat{\pmb{h}}_{\lambda}[\pmb{X}_n](\pmb{x}_1) -  \mathfrak{h}_{f}(\pmb{x}_1)| = O\left(\norm{\pmb{x}_1}\right)$. Then applying the Cauchy-Schwarz inequality, we get
\begin{align}\label{eq:a21}
 \mathbb{E}\left[ \left(\hat{\pmb{h}}_{\lambda}[\pmb{X}_n](\pmb{x}_1) - \mathfrak{h}_{f}(\pmb{x}_1)\right)^2  I\left\{\norm{\pmb{x}_1}_2 \le n^{\beta}\right\}\right] \lesssim \left\{n^{(K+ 3)\beta}\, \bar\Delta_{\lambda,n}\right\}^{1/2}.
\end{align}
Next, we consider the situation when $||\pmb x_i||_2$ is large.  Using assumption 2, it follows
\begin{align}\label{eq:a22}
	 \mathbb{E}\left[ \left(\hat{\pmb{h}}_{\lambda}[\pmb{X}_n](\pmb{x}_1) - \mathfrak{h}_{f}(\pmb{x}_1)\right)^2  I\left\{\norm{\pmb{x}_1}_2 > n^{\beta}\right\}\right] \lesssim n^{-\delta\beta}~.
\end{align}
Combining \eqref{eq:a21} and \eqref{eq:a22} gives the bound on $	\Delta^{(2)}_{\lambda,n}$ as 
\begin{align}\label{eq:a23}
	\Delta^{(2)}_{\lambda,n} \lesssim \left\{n^{(K+ 3)\beta}\, \bar\Delta_{\lambda,n}\right\}^{1/2} + n^{-\delta\beta}~.
\end{align}
Now, recall \eqref{eq:a7} in Lemma~\ref{Technical_lemma_1} upper bounds $\bar\Delta_{\lambda,n}$  by a function of $\Delta^{(2)}_{\lambda,n}$. Using  \eqref{eq:a7}  and \eqref{eq:a23}, we get 
\begin{align}\label{eq:a24}
		\Delta^{(2)}_{\lambda,n} \lesssim n^{(K+ 3)\beta/2}\left\{\, \lambda^{-(K+1)} \mathcal{S}_{\lambda}[\hat{\pmb{h}}_{\lambda,n+1}] + \lambda^2 \log n +    \lambda  (\log n)^{K+3}  \Delta_{\lambda,n}^{(2)} \right\}^{1/2} + n^{-\delta\beta}~.
\end{align}
Note that, $	\Delta^{(2)}_{\lambda,n}$ is bounded and so, Proposition~\ref{propLoss} implies $\mathcal{S}_{\lambda}[\hat{\pmb{h}}_{\lambda,n+1}]$ is bounded by $O(n^{-1})$. Finally, let $\lambda\asymp \Theta(n^{-\frac{1}{K+2}})$ and substitute $\beta = \frac{2}{(K+2)(K+3+\delta)}$ in \eqref{eq:a24} to obtain
\begin{align*}
		\Delta^{(2)}_{\lambda,n} \lesssim n^{-\frac{\delta}{(K+2)(K+3+2\delta)}}(\log n)^{K+3},
\end{align*}
which completes the proof of Theorem~2.
\section{Proof of Proposition 3}\label{proof:bandwidth}
First note that 
\begin{align*}
	&K_{\lambda} \left( (u_i, s_i); (u_j, s_j)   \right) / K_{\lambda} \left( (x_i, s_i); (x_j, s_j)   \right) = \exp\left\{-\frac{1}{2\lambda}(u_i -u_j)^2 + \frac{1}{2\lambda}(x_i-x_j)^2   \right\} \\=& \exp \left\{ -\frac{1}{2\lambda} \left[\alpha^2(\eta_i - \eta_j)^2 -   2\alpha (\eta_i - \eta_j)(x_i-x_j)   \right] \right\} := I_1.
\end{align*}
For any fixed $n$,  we have $x_{\max} - x_{\min} \le C_1$ and $\eta_{\max} - \eta_{\min} \le C_2$ for some quantities $C_1$ and $C_2$. Then the above is bounded by 
$$I_2:=\exp\left\{ -2^{-1}\lambda^{-1}\alpha(C_2^2 \alpha - 2C_1 C_2)   \right\}.
$$ 
The above ratio for $\nabla K_{\lambda}$ equals $I_1 {(u_i - u_j)}/{(x_i - x_j)}$, which is bounded in magnitude by $I_2\, {(C_3 + C_2)}/{C_3}$ where $C_3 = \min_{i \neq j} |x_i - x_j|$ and $C_3 > 0$ as the distribution of $Y$ in (1.1) is continuous.

Now consider the estimators
\begin{align*}
	\hat{\delta}_{\lambda, i}^{\text{IT}}(U, S) &= u_i + \sigma^2(1+\alpha^2) \hat g_i = y_i + \alpha \eta_i + \sigma^2(1+\alpha^2) \hat g_i, \text{ and, }\\
		\hat{\delta}_{\lambda, i}^{\text{IT}}(Y, S) &= y_i + \sigma^2(1+\alpha^2) \hat h_i~,
\end{align*}
where, for an arbitrary fixed value of $\lambda$, $\hat g_i$ and $\hat h_i$ are solutions from (2.4) using $(u, s)$ and $(y, s)$ respectively. Note that,
\begin{align*}
	\hat L_n(\lambda, \alpha) = \frac{1}{n} \sum_i \left( 	\hat{\delta}_{\lambda, i}^{\text{IT}}(U, S)  - v_i  \right)^2 - \sigma^2 (1+ \alpha^{-2}).
\end{align*}
Taking expectation and using the fact that $V$ is conditionally independent of $(U, S)$, we get,
\begin{align*}
	\E \{\hat L_n (\lambda, \alpha)\} = \E \left[    \frac{1}{n} \sum_i \left( 	\hat{\delta}_{\lambda, i}^{\text{IT}}(U, S)  - \theta_i  \right)^2   \right].
\end{align*}
For any fixed $n$,
\begin{align*}
	D_i := 	\hat{\delta}_{\lambda, i}^{\text{IT}}(U, S)  - \hat{\delta}_{\lambda, i}^{\text{IT}}(U, S) = \alpha \eta_i + \sigma^2\left[ (1+\alpha^2)\hat g_i - \hat h_i  \right].
\end{align*}
Now, if the optimization in (2.4) is strictly convex, then for any small $\alpha$, there exists $\epsilon_{\alpha}$ such that $\max_i |\hat g_i - \hat h_i| < \epsilon_{\alpha}$ and $\epsilon_{\alpha} \downarrow 0$ as $\alpha \downarrow 0$ and the result stated in this proposition follows.

%% file: append-real.tex
\section{Further details on simulation and Real Data Illustrations}\label{A.data}
\subsection{Simulation 2: Integrative estimation in two-sample inference of sparse means}\label{sec.simul.twosample}
This section considers compound estimation in two-sample inference. Let $X_{1i}$ and $X_{2i}$ be two Gaussian random variables. Denote $\mu_{1i}=\mathbb E(X_{1i})$ and $\mu_{2i}=\mathbb E(X_{2i})$, $1\leq i\leq n$. Suppose we are interested in estimating the differences $\pmb\theta=\{\mu_{1i}-\mu_{2i}: 1\leq i\leq n\}$. The primary statistic is given by $\pmb Y=\{X_{1i}-X_{2i}: 1\leq i\leq n\}$. However, it is argued in \cite{Caietal19} that the primary statistic $\pmb Y$ is \emph{not} a sufficient statistic. Consider the case where both $\pmb\mu_1$ and $\pmb\mu_2$ are individually sparse. Then an important fact is that the union support $\mathcal U=\{i: \mu_{1i}\neq 0 \mbox{ or } \mu_{2i}\neq 0\}$ is also sparse. The intuition is that the sparsity structure of $\pmb\theta$ is captured by an auxiliary parameter $\pmb\eta=\{\mu_{1i}+\mu_{2i}: 1\leq i\leq n\}$. Our idea is to construct an auxiliary sequence $\pmb S=\{X_{1i}+X_{2i}: 1\leq i\leq n\}$ and incorporate $\pmb S$ into inference to improve the efficiency\footnote{It can be shown that $\{(X_{1i}-X_{2i}, X_{1i}+X_{2i}): 1\leq i\leq n\}$ is minimal sufficient and retains all information about $\pmb\theta$.}. 

To illustrate the effectiveness of the integrative estimation strategy, we simulate data according to the following two settings and obtain primary and auxiliary data as  $\pmb Y=\{X_{1i}-X_{2i}: 1\leq i\leq n$ and $\pmb S=\{X_{1i}+X_{2i}: 1\leq i \leq n\}$. 
\begin{description}
\item Setting 1: $X_{1i}$ and $X_{2i}$ are generated from $
X_{1i} \sim \cN(\mu_{1i}, 1)$ and $X_{2i} \sim \cN(\mu_{2i}, 1)$, where 
\begin{eqnarray*}
	\pmb\mu_1[1:k] = 2.5, &\qquad \pmb\mu_2[1:k] = 1\\
	\pmb\mu_1[k+1:2k] = 1, &\qquad \pmb\mu_2[k+1:2k] = 1\\
	\pmb\mu_1[2k+1:n] = 0, &\qquad \pmb\mu_2[2k+1:n] = 0
\end{eqnarray*}		
The sparsity level of $\pmb\theta$ is controlled by $k$. We fix $n=1000$ and vary $k$ from $50$ to $450$ to investigate the impact of sparsity level on the efficiency of different methods. 
\item Setting 2: $X_{1i}$ and $X_{2i}$ are generated from $
X_{1i} \sim \cN(\mu_{1i}, 1)$ and $X_{2i} \sim \cN(\mu_{2i}, 1)$, where 
\begin{align*}
	\pmb\mu_1[1:k] = 1, &\qquad \pmb\mu_2[1:k] = 1\\
	\pmb\mu_1[k+1:500] = 2.5,  &\qquad \pmb\mu_2[k+1:500] = 1\\
	\pmb\mu_1[501:n] = 0, &\qquad \pmb\mu_2[501:n] = 0
\end{align*}			
The primary parameter $\pmb\theta$ becomes more sparse when $k$ increases. We fix $n=1000$ and vary $k$ from $50$ to $450$ to investigate the efficiency gain of NIT.  
\end{description}
\begin{figure*}
\begin{subfigure}{0.47\textwidth}
\centering
\includegraphics[width=1\textwidth]{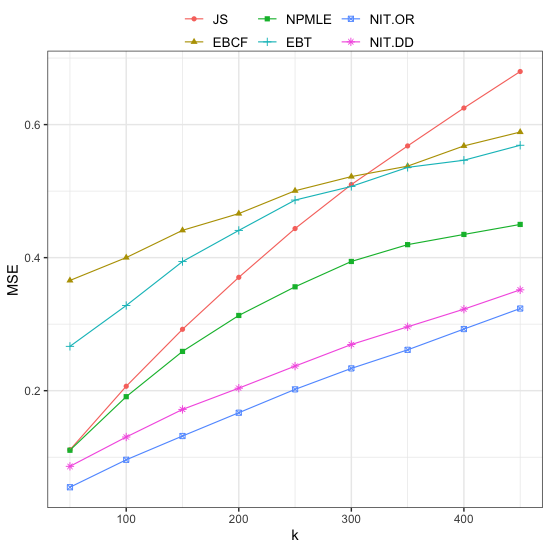}
\caption{Setting 1}
\label{fig:setting1}
\end{subfigure}
\hfill
\begin{subfigure}{0.47\textwidth}
\centering
\includegraphics[width=1\textwidth]{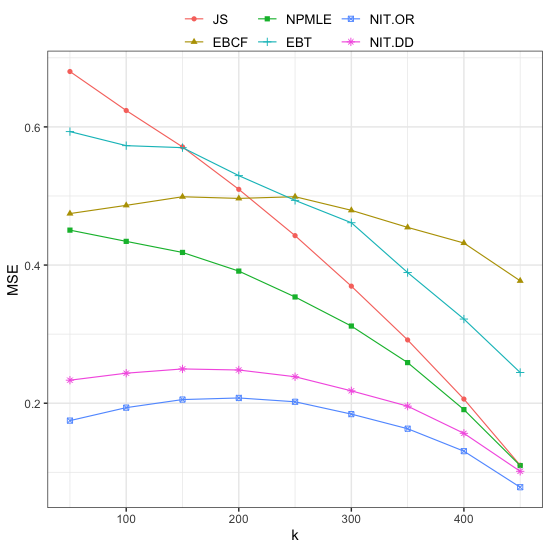}
\caption{Setting 2}
\label{setting2}
\end{subfigure}
\hfill
\caption{Two-sample inference of sparse means.}
\label{two_sample_homo}
\end{figure*}	
We apply different methods to simulated data and calculate the MSEs using $100$ replications. The simulation results are displayed in Figure \ref{two_sample_homo}. The following can be observed.
\begin{itemize}
\item [(a).] The side information provided by the auxiliary sequence can be highly informative for reducing the estimation risk. Our proposed methods (NIT.DD, NIT.OR) have smaller MSEs than competing methods (EBCF, JS, NPMLE, EBT). The efficiency gain over univariate methods (JS, EBT, NPMLE) is more pronounced when signals become more sparse.   
\item [(b).] EBCF is dominated by NIT, and can be inferior to univariate shrinkage methods. 
\item [(c).] The class of linear estimators is inefficient under the sparse setting. For example, the NPMLE method dominates the JS estimator, and the efficiency gain increases when the signals become more sparse. 
\end{itemize}	
\subsection{Gene Expressions Estimation example}\label{A.data1}
The data considered in this analysis was collected in \cite{sen2018distinctive} via RNA sequencing. The set of genes in the sequencing kit was same across all the experiments. 
The standard deviations of the expressions values corresponding to the different genes were estimated from related gene expression samples which contain replications under different experimental conditions.  Pooling data across these experiments, unexpressed and lowly expressed genes were filtered out. The resultant data consist of around 30\% of the genes. We consider the estimation of the mean expression levels of $n=3000$ genes. The primary parameter $\pmb{\theta}$ is estimated based on primary vector $\pmb{Y}$ and two auxiliary sequences $\pmb{S}_{\sf U}$ and $\pmb{S}_{\sf I}$. 

\begin{figure}[!t]
\centering	
\includegraphics[width=0.8\textwidth]{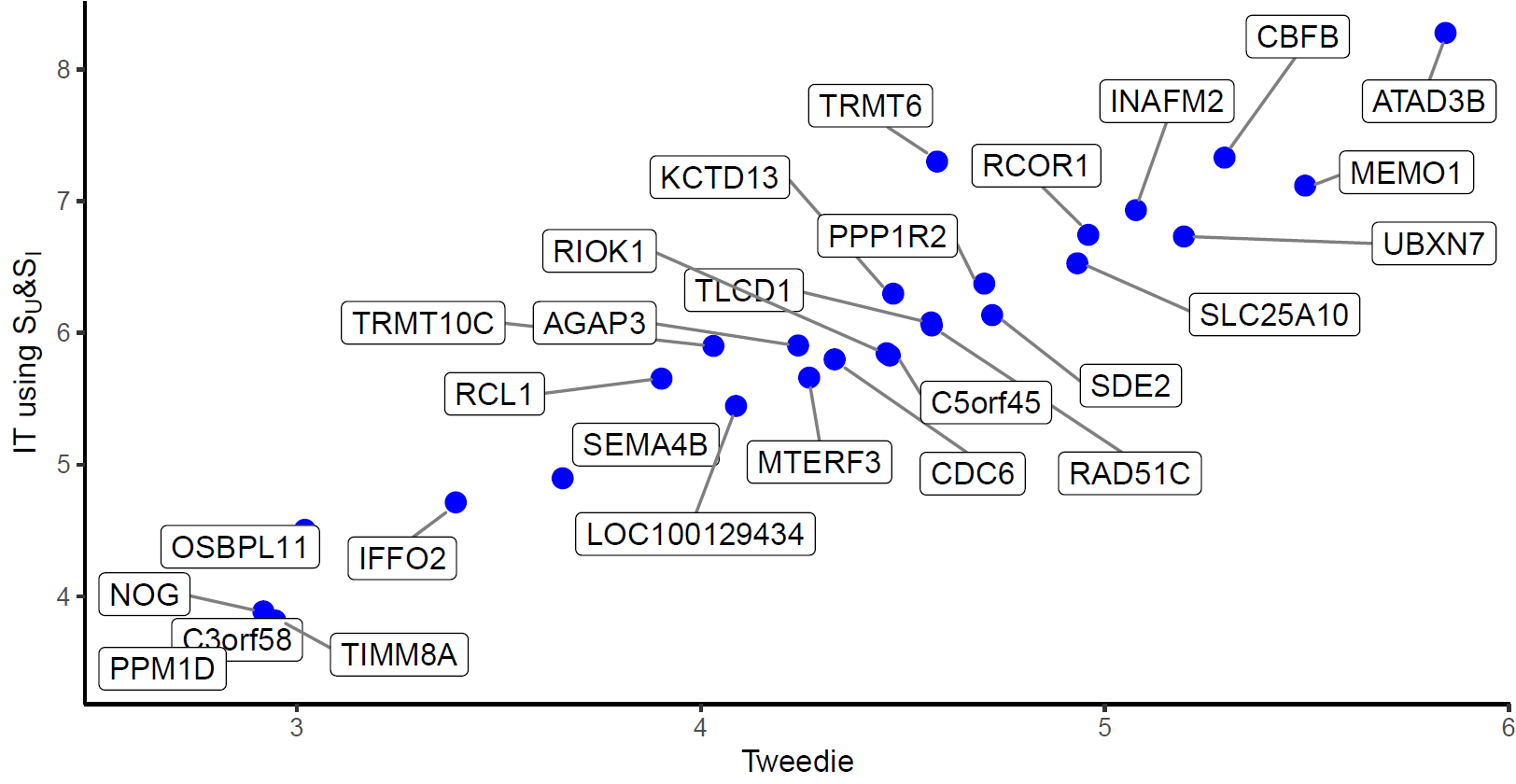}
\includegraphics[width=0.8\textwidth]{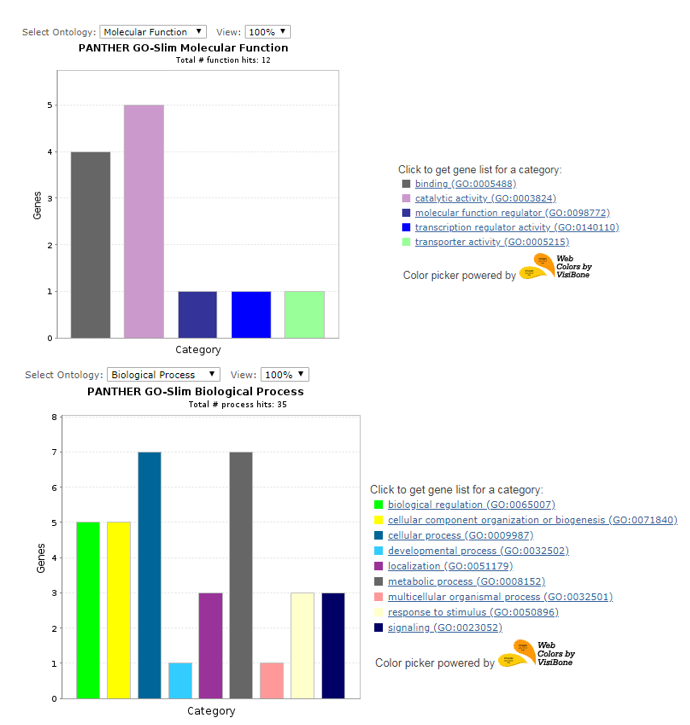}
\caption{Top panel: Scatterplot and names of genes where Tweedie and Integrated Tweedie effect size estimates disagreed by more than 50\%. The other panels show the different molecular function types and biological processes that are impacted by these genes.}\label{fig.A.gene}
\end{figure}

In Figure~\ref{fig.A.gene} (top panel), we list the $28$ genes for which the Tweedie and integrative Tweedie estimates disagree by more than 50\%. 
According to PANTHER (Protein ANalysis THrough Evolutionary Relationships) Classification System \citep{mi2012panther}, those genes impact $12$ molecular functions and $35$ biological processes in human cells. The bottom two panels of Figure~\ref{fig.A.gene} present the different function and process types that are impacted. 
\subsection{Leveraging auxiliary information in predicting monthly sales}
We consider the total monthly sales of beers across $n= 866$ stores of a retail grocery chain. These stores are spread across different states in the USA as shown in Figure~\ref{us_map}. The data is extracted from \cite{Broetal08}, which has been widely studied for inventory management and consumer preference analyses; see also \cite{Broetal12} and the references therein. 

Let $\bm{Y}^{t}$ be the $n$ dimensional vector denoting the monthly sales of beer across the $n$ stores in month $t \in \{1,\ldots,12\}$. For inventory planning, it is economically important to estimate future demand. In this context, we consider estimating the monthly demand vector (across stores) for month $t$ using the previous month's sales $\bm{Y}^{t-1}$. 
We use the first six months $t=1,\ldots,6$ for  estimating store demand variabilities  $\hat{\sigma}^2_i, i=1,\ldots,n$. For $t=7,\ldots,12$, using estimators based on month $t$'s sales, we calculate their demand prediction error for month $t+1$ by using its monthly sale data for validation. Among the estimators, we consider the modified James-Stein (JS) estimator of \citet{Xieetal12}: 
$$\hat{\bm{\theta}}_i^{t+1}[\textsf{JS}] = \widehat{\textsf{JS}}_i^t +\bigg[1-\frac{n-3}{\sum_i \hat \sigma_i^{-2} (Y_i^t-\widehat{\textsf{JS}}_i^t)^2}\bigg]_+(Y_i^t - \widehat{\textsf{JS}}_i^t) \text{ where } \widehat{\textsf{JS}}_i^t = \frac{\sum_{i=1}^n\sigma_i^{-2} Y_i^t }{\sum_{i=1}^n \sigma_i^{-2}},$$
as well as the Tweedie (T) estimator $\hat{\bm{\theta}}_i^{t+1}[\textsf{T}]= Y_i^t +\hat \sigma_i \hat h_i$ where $\hat h_i$ are estimates of $ \nabla_1 \log f(\hat \sigma_i^{-1} Y_i^t )$ based on the marginal density of standardized sales. We also consider the sales of three other products: milk, deodorant and hotdog from these stores. They are not directly related to the sale of beers but they might contain possibly useful information regarding consumer preferences to beers particularly as they share zip-code and other store specific responses. We use them as auxiliary sequences in our NIT methodology. 
\begin{figure}[!h]
	\includegraphics[width=1\textwidth]{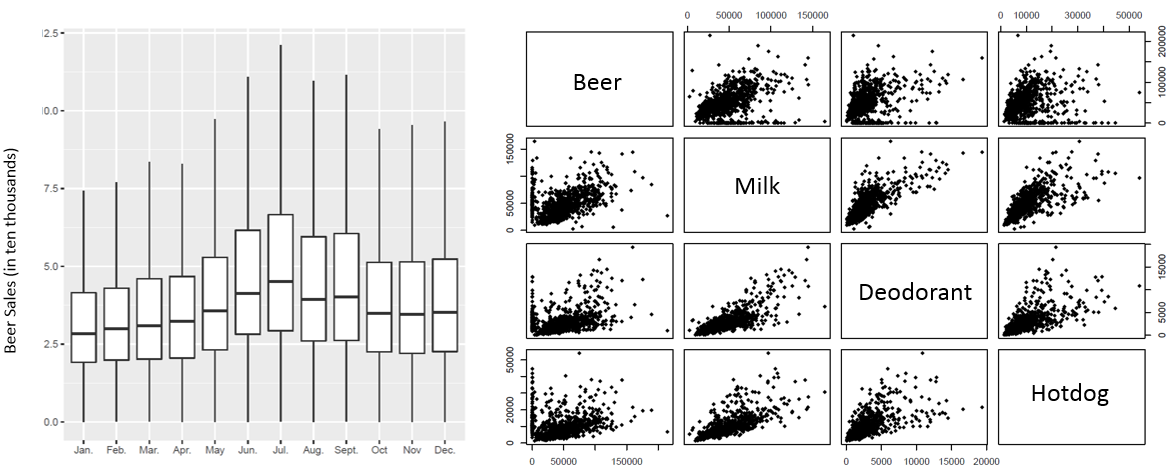}
	\caption{ Distribution of monthly sales of beer across stores (on left) and the pairwise distribution of joint sales of different products in the month of July (in right).}
	\label{fig:groceries-1}
\end{figure}	
Figure~\ref{fig:groceries-1} shows the distribution of beer sales (across stores) for different months and the pairwise distribution of the sales of different products. 
Further details about the dataset is provided in Section \ref{A.data2}.

In Table \ref{tab:groceries-1}, we report the  average \% gain in predictive error by the James-Stein (JS), Tweedie (T) and integrative Tweedie (IT) estimators (using different combinations of auxiliary sequences) over the naive estimator $\hat{\delta}^{t+1,\text{naive}}=\bm{Y}^t$ for the demand prediction problem at $t=7,\ldots,12$. 
Using auxiliary variables via our proposed NIT framework yields significant additional gains over non-integrative methods. However, the improvement slackens as an increasing number of auxiliary sequences are incorporated. It is to be noted that the demand data set is highly complex and heterogeneous and $n=866$ may not be adequately large for conducting successful non-parametric estimation. Hence suitably anchored parametric JS estimator produces better prediction than nonparametric Tweedie. Also, as demonstrated in Table \ref{tab:groceries-3}, there are months where shrinkage estimation methods do not yield positive gains. Nonetheless, the NIT estimator produces significant advantages over competing methods. It produces on average 7.7\% gain over unshrunken methods and attains an additional 3.7\% gain over non-parametric shrinkage methods.   
\begin{table}[!h]
	\centering
	\caption{Average \% gains over the naive unshrunken estimator for monthly beer sales prediction}
	\scalebox{0.9}{\begin{tabular}{|c|c|c|c|c|c|c|c|c|}
			\hline
			JS    & Tweedie & IT-Milk & IT-Deodorant & IT-Hotdog & IT-M\&D & IT-M\&H & IT-D\&H & IT-M\&D\&H \bigstrut\\
			\hline
			5.7   & 4.0     & 6.0     & 7.1   & 6.8   & 6.1   & 6.6   & 7.5   & 7.7 \bigstrut\\
			\hline
		\end{tabular}%
	}
	\label{tab:groceries-1}%
\end{table}%
\subsection{Additional details on the monthly sales data example}\label{A.data2}
\begin{figure}[!h]
\centering
\includegraphics[trim=2cm 3cm 0cm 3cm, clip=true, width=0.85\textwidth]{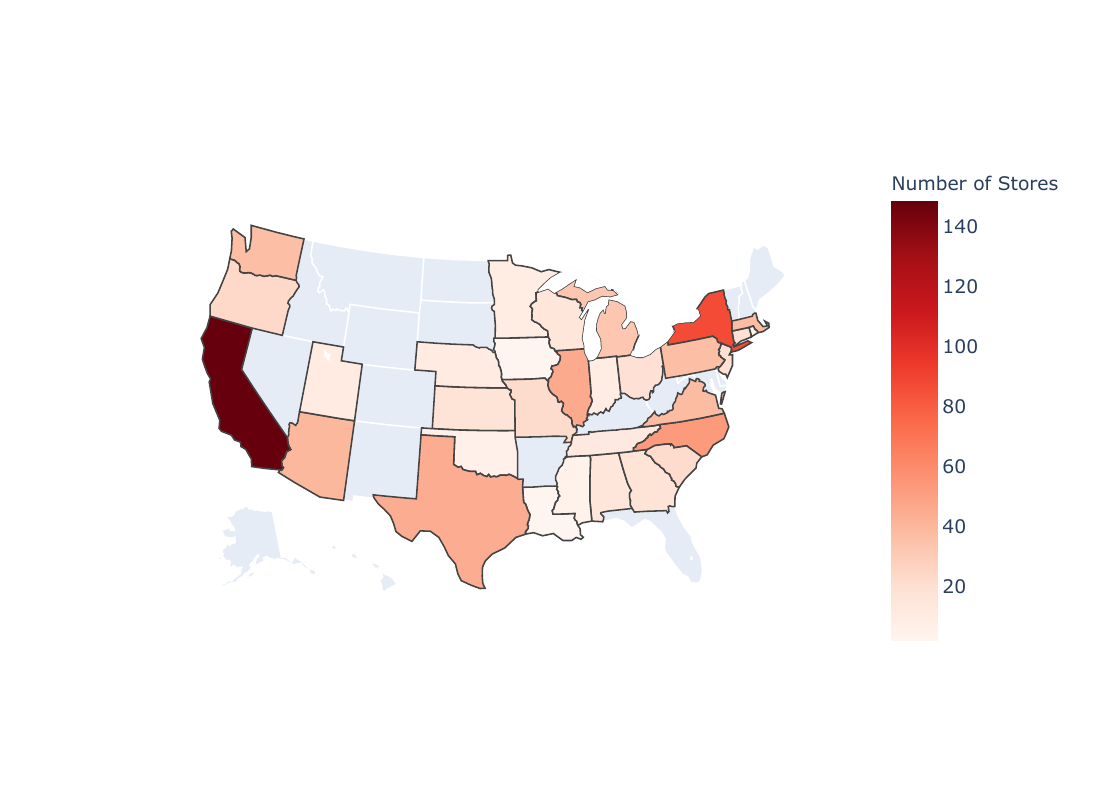}
\caption{Distribution of the 866 stores across different states in USA.}\label{us_map}
\end{figure}

Here, we consider the monthly sales at the store level for $3$ additional commodities: milk, deodorant and hotdog.  The distribution of $866$ store across different US states is shown in Figure~\ref{us_map} and Table~\ref{tab:correlation_matrix} shows the correlation between the different products. 

\begin{figure}[!t]
\centering
\includegraphics[width=1\textwidth]{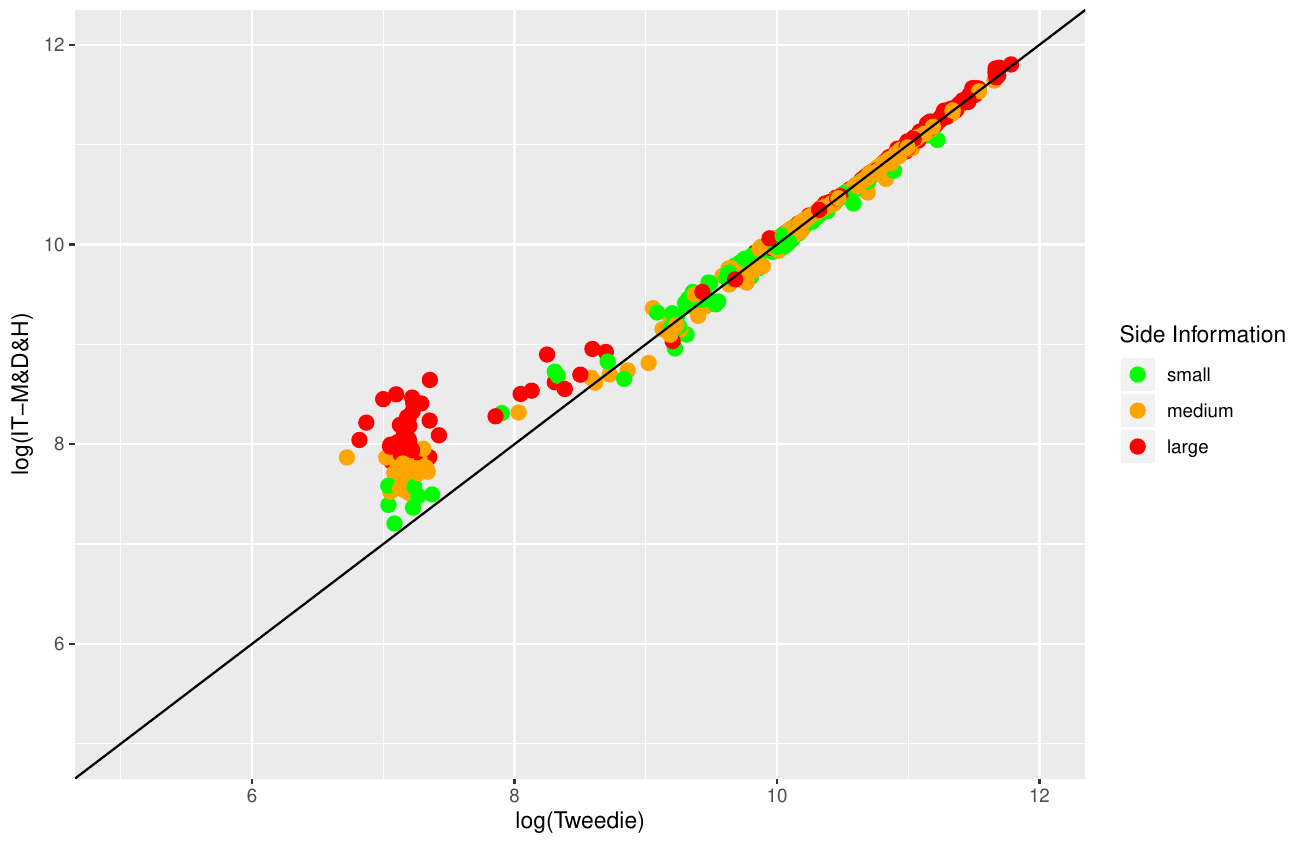}
\caption{Scatterplot of the logarithm of beer demand estimates in the month of August. The magnitudes of the corresponding auxiliary variables used in the IT estimate are reflected in the different colors. We can see that the most significant differences between Tweedie and integrative Tweedie are observed in the left-tails. This shows the region where the side information is most helpful. }
\label{fig:pred_compare}
\end{figure}

In Table \ref{tab:groceries-3}, we report the  average \% gain in  predictive error by the JS, T and IT estimators (using different combinations of auxiliary sequences) over the naive unshrunken estimator $\hat{\delta}^{t,\text{naive}}=\pmb{Y}^{t-1}$ for the demand prediction problem at $t=7,\ldots,12$. 
For estimator $\hat{\pmb{\delta}}$ we report, 
\begin{align*}
\mathsf{Gain}_t( \hat{\pmb{\delta}}) = \frac{\sum_{i=1}^n \hat{\sigma}_i^2(\hat{\delta}_i^t - \hat{y}^{t}_i  )^2 }{\sum_{i=1}^n \hat{\sigma}_i^2(\hat{\delta}^{t,\text{naive}} - \hat{y}^{t}_i  )^2 } \times 100 \% \quad \text{ for } \;\; t=7,\ldots, 12. 
\end{align*}
The last column in Table \ref{tab:groceries-3} reports the average performance of these methods over the six successive trails. 
\begin{table}[!h]
\centering
\caption{Correlation matrix of the monthly sales of different products.}
\begin{tabular}{*7c}
\hline
\multicolumn{2}{@{}c}{Products} &
{(1)} & {(2)} & {(3)} & {(4)} \\    
\hline
(1)  & Beer&1.00\\
(2)  & Milk &0.33&1.00\\
(3)  & Deod &0.16&0.63&1.00\\
(4)  & Hotdog &0.84&0.38&0.19&1.00\\
\hline
\end{tabular}
\label{tab:correlation_matrix}
\end{table}
\begin{table}[!h]
\centering
\caption{Month-wise \% gains for monthly beer sales prediction over the naive unshrunken estimator.}
\scalebox{0.8}{\begin{tabular}{|l|c|c|c|c|c|c|c|}
\hline
& \textbf{July} & \textbf{August} & \textbf{September} & \textbf{October} & \textbf{November} & \textbf{December} & \textbf{Average} \bigstrut\\
\hline
\textbf{James-Stein} & 9.7   & 2.4   & 10.8  & -2.7  & -16.2 & -3.7  & 5.7 \bigstrut\\
\hline
\textbf{Tweedie} & 7.5   & 7.5   & 9.6   & -7.2  & -22.6 & -2.8  & 4 \bigstrut\\
\hline
\textbf{IT -Milk} & 11.7  & 5.2   & 9.4   & -7.4  & -8.8  & -8.2  & 6 \bigstrut\\
\hline
\textbf{IT -Deo} & 11.3  & 5.1   & 10.7  & -10.6 & -13.7 & 3.7   & 7.1 \bigstrut\\
\hline
\textbf{IT -Hotdog} & 12.4  & 2.6   & 11.9  & -3.2  & -13.2 & -6.5  & 6.8 \bigstrut\\
\hline
\textbf{IT-M\&D} & 10.7  & 5.9   & 9.8   & -7.4  & -8.7  & -7    & 6.1 \bigstrut\\
\hline
\textbf{IT-M\&H} & 10.3  & 5.7   & 10.8  & -4.3  & -10.3 & -4.8  & 6.6 \bigstrut\\
\hline
\textbf{IT-D\&H} & 11.7  & 6.8   & 11    & -8.2  & -9.1  & -0.6  & 7.5 \bigstrut\\
\hline
\textbf{IT-M\&D\&H} & 11.2  & 6.8   & 10.9  & -8.1  & -7.2  & 1.8   & 7.7 \bigstrut\\
\hline
\end{tabular}}%
\label{tab:groceries-3}%
\end{table}%
In Figure \ref{fig:pred_compare}, we compare the prediction of monthly sales in August using Tweedie and IT-M\&D\&H. The magnitude of side-information is marked using different colors. We can see that the most significant differences between Tweedie and integrative Tweedie are observed in the left-tails. 
